\documentclass[11pt,a4paper]{article}
\pdfoutput=1
\usepackage{jheppub}
\usepackage{amsfonts}
\usepackage{bbm}
\usepackage{graphicx}
\usepackage{caption}
\usepackage{subcaption}
\usepackage{bigints}
\usepackage[normalem]{ulem}
\usepackage{slashed}
\usepackage{subcaption}

\def\h{\eta}

\def\h{\eta}

\newcommand{\RN}[1]{%
  \textup{\uppercase\expandafter{\romannumeral#1}}%
}

\def\bea{\begin{eqnarray}}
\def\eea{\end{eqnarray}}
\def\be{\begin{equation}}
\def\ee{\end{equation}}
\def\ba{\begin{align}}
\def\ea{\end{align}}

\newcommand{\bem}{\begin{pmatrix}}
\newcommand{\eem}{\end{pmatrix}}

\def\sl{$SL(2,R)$}

\def\={\;  = \;}
\def\+{\, + \,}

\def\bar{\overline}

\def\rt2{\sqrt{2}}

\title{Quantum information scrambling and quantum chaos in little string theory}

\author[a]{\small{Sandip Mahish and Karunava Sil}}
\affiliation[a]{\small{School of Basic Sciences, Indian Institute of Technology Bhubaneswar, Bhubaneswar 752050, India}}

\emailAdd{sm19@iitbbs.ac.in}
\emailAdd{ks45@iitbbs.ac.in}

\abstract{In the current manuscript we perform a systematic investigation about the effects of nonlocal interaction to the spread of quantum information in many body system. In particular, we have studied how nonlocality influence the existing bound on the growth rate of the commutator involving two local operators, the butterfly velocity. For this purpose, we consider the nonlocal theory on the worldvolume of $N\gg 1$, NS$5$ branes arising in the limit of vanishing string coupling, the `little string theory'. A direct evidence of nonlocality can be realized from the `volume law' behavior for the most dominant part of holographic entanglement entropy. We obtain the butterfly velocity by studying the dynamics of the near horizon geometry backreacted by a high energy quanta in the form of a shockwave resulting from an early perturbation on the corresponding thermofield double state. We observe that the butterfly velocity increases with the nonlocal scale of little string theory, the inverse Hagedorn temperature $\beta_{h}$, indicating a faster rate of information spread due to the nonlocal interaction. The same conclusion follows as the disruption of two sided mutual information is observed to occur at a faster rate for higher values of $\beta_{h}$. Finally, we realize a direct connection between the parameters of quantum chaos and the quasinormal modes for collective excitations through the phenomenon of `pole skipping'.}

\makeatletter
\gdef\@fpheader{}
\makeatother

\begin{document}

%

\maketitle

\section{Introduction}
Recently, there has been a growing interest in quantifying the nature of quantum information spread in strongly interacting many body system. In quantum many body system, the spread of information effectively depends on the nature of commutator between two local generic operators. In particular, the commutator behaves differently when evaluated in chaotic and integrable systems. There exists fundamental bounds on such commutators as evaluated for different quantum theories which in turn reflects how fast quantum information can scramble from being accessible to a few local degrees of freedom to the entire system.\footnote{We appreciate the nicely written lecture notes \cite{BS} which helps us to understand the basic idea and also recent developments regarding the quantum information scrambling in many body quantum theory.}

As an example, a relativistic system that is governed by the rules of Lorentz symmetry does not allow any correlation between two space like separated local operators. In this case causality sets the bound to the information spread by the velocity of light. In a non-relativistic system, on the other hand two local operators can have finite overlap even if they are largely separated in space as compared to the elapsed time. However, a bound on the information propagation still exists for the non-relativistic system, known as Lieb-Robinson bound as observed in a discrete lattice system \cite{Lieb:1972wy}. Here the corresponding `Lieb-Robinson velocity' defines a bound for the growth of operator norm of two Heisenberg operators $W(t,x)$ and $V(t,x)$. Lieb-Robinson bound depends on the details of the UV physics and hence it is independent of the state of the system. However, at lower energy scale such state independent bound is not relevant. There exists much tighter and slower speed limit, for example the butterfly velocity \cite{Shenker:2013pqa, Roberts:2016wdl} that bounds the propagation of all kinds of physical excitations which are important at the low energy density or temperature. Butterfly velocity is a state dependent quantity as it is given in terms of the IR scale of the underlying theory. Instead of operator norm butterfly velocity is governed by the growth of commutator square involving generic operators. The butterfly velocity in general local quantum theory follows several bounds and inequalities. In this paper we will particularly be interested in how these bounds and inequalities gets modified due to the inherent non locality that the little string theory possess. Moreover, we will mainly focus on the connection between information scrambling and quantum chaos for a quantum many body theory with non-local interaction.

Chaotic behavior in a quantum system with large number of degrees of freedoms can be quantified as the rate with which quantum information spreads across the entire system. Moreover, in many body systems which are governed by complicated hamiltonian, information encoded within any small subsystem scrambles among the degrees of freedom of the whole system such that it cannot be recovered by doing local measurement on that particular subsystem. Let the system under consideration is defined initially by the density matrix $\rho_{0}$ which evolves in time according to the usual time evaluation rules in Heisenberg picture, $\rho(t)=e^{-iHt}\rho_{0}e^{iHt}$, $H$ being the hamiltonian of the system. If the system is partitioned into two parts, $A$ and $B$ then any local measurement on subsystem $A$ can be performed with the reduced density matrix $\rho_{A}=tr_{B}(\rho(t))$. As time evolves the subsystem $A$ starts loosing information and it is no longer defined by the reduced density matrix $\rho_{A}$ but instead by the thermal ensemble density matrix $\rho_{\beta}$ at some inverse temperature $\beta$ and the information about the initial state of $A$ is no longer available on local measurements. Typically quantum chaos is characterized by the thermal average of the following double commutator for two generic operator $W$ and $V$ as \cite{larkin, Shenker:2013pqa},
\begin{equation}\label{OTOC}
C(t,x)=-<[W(t,x),V(0)]^2>_{\beta}.
\end{equation}
In the above equation the double commutator can be simplified only to observe that it is proportional to the out of time ordered product of operators $W$ and $V$, the OTOC. In quantum many body systems, the chaotic behavior is captured by the exponential growth of the OTOC, $C(t,x)$ in the above equation given as, \cite{Roberts:2014isa, Shenker:2014cwa, Kitaev:2017awl, Polchinski:2015cea, Polchinski:2016xgd, Maldacena:2016hyu, Maldacena:2016upp, Jensen:2016pah, Gu:2016oyy, Patel:2016wdy, Alishahiha:2016cjk}
\begin{equation}\label{OTOC2}
C(t,x)=\frac{1}{N^2}e^{\lambda_{L}\left(t-\frac{|x|}{v_{b}}\right)},
\end{equation}
where the overall growth rate is determined by the Lyapunov exponent $\lambda_{L}$. As shown in \cite{Maldacena:2015waa}, that there exists an upper bound to the Lyapunov exponent given in natural unit as $\lambda_{L}\le \frac{2\pi}{\beta}$ which saturates for maximally chaotic systems. More specifically, for a large class of holographic theories with black hole in Einstein's gravity this bound is observed to be saturated \cite{Grozdanov:2017ajz, Blake:2018leo, Grozdanov:2018kkt} leading to the conclusion that black holes are the fastest scramblers of quantum information \cite{Sekino:2008he},\cite{Lashkari:2011yi}.  In equation (\ref{OTOC}), the commutator measures the correlation between a small perturbation $V$ at some early time and some other operator $W$ inserted at position $x$ and time $t$. The growth of this early perturbation is characterized by the butterfly velocity $v_{b}$ which defines an effective light cone such that the operator $W$ if inserted outside of it will not be effected by the early perturbation, $C(t,x)\approx 0$. However inside the light cone the commutator is finite and as the initial perturbation scrambles with time into the degrees of freedom of the entire system, $C(t,x)$ also grows until some time scale $t_{\ast}$, known as the scrambling time at which $C(t,x)$ becomes $\mathcal{O}(1)$ \cite{Hayden:2007cs}.

There is another important interpretation of butterfly velocity in the context of AdS/CFT correspondence. It determines the causal structure of the bulk geometry \cite{Qi:2017ttv} which unlike the microcausality in CFT, is not that straight forward, especially near the black hole horizon. Given a region $A$ on the boundary, it's corresponding entanglement wedge is defined as the bulk region which is bounded by the region $A$ itself and the Ryu-Takayanagi (RT) surface that is homologous to $A$. Now consider a particle inside the bulk which is entangled with some reference state. With evolving time as this particle falls deep inside the bulk, its entanglement with the reference state can only be recovered as long as it stays inside the entanglement wedge. So the region $A$ must grow with a rate such that the corresponding RT surface also penetrate sufficiently deep into the bulk to be able to confine the particle inside the entanglement wedge. The growth rate of $A$ on the boundary is exactly equal to the butterfly velocity $v_{b}$. The fact that it is not possible for the region $A$ to grow any faster than the butterfly velocity $v_{b}$, sets a restriction for the particle speed in the bulk. This is how the bulk causality can be interpreted from the knowledge of butterfly velocity on the boundary. In \cite{Mezei:2016wfz}, butterfly velocity was computed holographically for asymptotically AdS-BH spacetime in general dimension to get,
\begin{equation}
v_{b}=\sqrt{\frac{d+1}{2d}},
\end{equation}
where $d$ is the spatial dimension of the boundary field theory. Usually in local quantum theory the butterfly velocity is bounded by the speed of light. However, in the presence of nonlocal interactions this bound can be violated. As an example in \cite{Fischler:2018kwt}, the authors observed that in $\mathcal{N}=4$ non-commutative SYM theory, $v_{b}$ happens to depend on the non-commutative parameter $\theta$ and for sufficiently large $\theta$ it can even exceed the speed of light. This violation is a result of the explicit breakdown of Lorentz invariance due to the non-commutative parameter.

The connection between quantum chaos and information scrambling can be better understood in a dynamical setting as provided by the process of thermalization where the expectation value of any simple observable decays with time to the thermal value. In other words, as a system thermalize, it looses the memory of initial state resulting in the generation of effective entropy for entanglement. There exists a number of interesting papers on the characterization of entanglement growth in free field theory \cite{Calabrese:2005in, Calabrese:2007rg, Cotler:2016acd}. In the presence of strong coupling, the simple picture of entanglement spreading by the flow of non-interacting particles in a free theory is no longer applicable. To deal with strong coupling the authors in \cite{Liu:2013iza, Liu:2013qca, MohammadiMozaffar:2018vmk} considered a set up which consists of injecting at $t=0$, an uniform energy density (global quench) for a negligible period of time and then letting the resulting excited state thermalize. It is observed that in $(1+1)$ dimensional quantum system as well as in higher dimensional holographic system the entanglement of any subsystem $\Sigma$ grows linearly in time,
\begin{equation}\label{ev}
\frac{dS_{\Sigma}(t)}{dt}=v_{E}s_{th}A_{\Sigma},
\end{equation}
with $s_{th}$ is the thermal entropy and $A_{\Sigma}$ is the area of the entangling subsystem $\Sigma$. Physically this growth of entanglement can be realized in terms of `entanglement tsunami' wave propagating inwards from the boundary of $\Sigma$ with velocity $v_{E}$ such that only the region which is already covered by the wave is entangled with the complimentary region $\Sigma^{c}$. In \cite{Calabrese:2005in}, $v_{E}$ was obtained for $(1+1)$ dimensional system to be equal to $1$ in natural unit. For higher dimensional holographic system with $d\ge 3$, the authors in \cite{Hartman:2013qma, Liu:2013iza, Liu:2013qca} calculated the following general result for $v_{E}$,
\begin{equation}
v^{holo}_{E}=\frac{\left(\eta-1\right)^\frac{\eta-1}{2}}{\eta^{\eta/2}},~~~~~~\eta=\frac{2d-2}{d}.
\end{equation}
Further, in \cite{Casini:2015zua}, it was shown that for relativistic theories in any dimension, $v_{E}$ cannot exceed the velocity of light, $v_{E}\le 1$. Finally, the authors in \cite{Mezei:2016wfz} managed to came up with the proof that for any unitary quantum theory, entanglement velocity must be bounded by the corresponding butterfly velocity,
\begin{equation}
v_{E}\le v_{b}.
\end{equation}
Later, in \cite{Mezei:2016zxg}, the same inequality was observed to hold for holographic theories as well.

In this paper we will concentrate on the holographic calculation of entanglement velocity by measuring the disruption of mutual information between two completely disjoint quantum theories coupled together forming a thermofield double state (TFD). For schematic conveniences, let us call the two copies of the quantum theories as $QFT_{L}$ and $QFT_{R}$, with the suffix denoting `Left' and `Right' respectively. A TFD state is defined on the tensor product of the Hilbert spaces of $QFT_{L}$ and $QFT_{R}$ such that given the energy eigenstates $|E_{i}>_{L}$ and $|E_{i}>_{R}$ of the two QFTs, it is the unique pure entangled state given as,
\begin{equation}
|TFD>=\frac{1}{Z^{1/2}}\sum_{i}e^{-\frac{\beta}{2}E_{i}}|E_{i}>_{L}|E_{i}>_{R}.
\end{equation}
\begin{figure}[h]
\centering
\begin{tabular}{c}
\includegraphics[width=.55\textwidth]{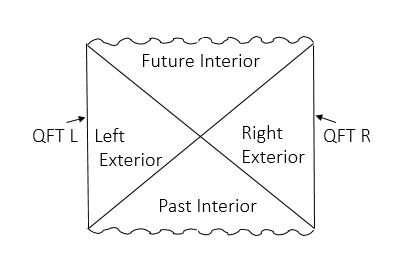}
\end{tabular}
\caption{Kruskal diagram for two sided black hole geometry}
\label{TFD}
\end{figure}
It is important to note that each of the two states on the left and right side are exactly thermal and the thermality in each copy arises due to the entanglement with the other one. In the language of AdS/CFT correspondence \cite{Maldacena:1997re}, thermofield double state is dual to the eternal black hole geometry which can be realized in Penrose diagram (see figure-\textbf{\ref{TFD}}) as a maximal analytic extension of the usual one sided black hole geometry \cite{Maldacena:2001kr}. Here the interior of the two black holes are connected through the wormhole geometry that indicates the entanglement between the two black holes.
Mutual information $I(A,B)$ between two subsystems $A\in QFT_{L}$ and $B\in QFT_{R}$ is defined as,
\begin{equation}\label{mi}
I(A,B)=S(A)+S(B)-S(A\cup B),
\end{equation}
where $S(A\cup B)$ is the entanglement entropy of the union of $A$ and $B$. Now initially there is a large amount of entanglement between the two subregions as evident form the existing upper bound on the correlation between two operator $\mathcal{O}_{A}$ and $\mathcal{O}_{B}$ defined on $A$ and $B$ respectively given as \cite{Wolf:2007tdq},
\begin{equation}
I(A,B)\ge \frac{\left(<\mathcal{O}_A \mathcal{O}_B>-<\mathcal{O}_{A}><\mathcal{O}_{B}>\right)^2}{2<\mathcal{O}_{A}^2><\mathcal{O}_{B}^2>}
\end{equation}
In order to characterize the spread of entanglement we first need to disrupt the TDF structure by applying a small perturbation $W$ on the left side (say) at some early time $t_{i}$. In the bulk, this corresponds to some localized energy density injected near the boundary in the distant past. Due to the strong gravitational force, the excitation will accelerate with time and it's energy gets blue shifted as it reaches near the black hole horizon. This results in a shockwave modified geometry for the bulk spacetime. It is important to note that only the left-right correlation is destroyed in this case making the mutual information $I(A,B)$ to vanish for a sufficiently early time perturbation. As a result in equation (\ref{mi}), only $S(A\cup B)$ shows increasing behavior with time exactly the same way as in equation (\ref{ev}). That is the growth rate is controlled by the entanglement velocity $v_{E}$. The entanglement entropy $S(A)$ and $S(B)$ being unchanged, results $I(A,B)$ to vanish consequently.

As already mentioned the primary goal of this study is to investigate the effect of nonlocality to the spread of quantum information. In the context of black hole information paradox, quantum nonlocality might possibly be the fundamental ingredient towards the resolution of the paradox. In \cite{Giddings:2006sj}, it was argued that in a situation of strong gravitational interaction the dynamics is governed by the nonlocal degrees of freedom. These observations/findigs motivate us to study the dynamics of nonlocal information spreading in a particular quantum theory with inherent nonlocality, the little string theory \cite{DK}. Little string theory arises in the decoupling limit (vanishing string coupling $g_{s}$) of magnetically charged $N\gg 1$, NS$5$-branes in type-II string theory, $N$ being the number of five branes. In this case the dynamics of the NS5-branes decouples from the bulk even with finite string length $l_{s}=\sqrt{\alpha^{\prime}}$. The resulting theory on five branes is known as the little string theory which is a nonlocal theory without gravity, it exhibits T-duality upon compactification. Unlike the usual local QFT, the fundamental object in this theory are one dimensional strings. The impurity in the form of NS$5$-branes breaks the original $SO(9,1)$ Lorentz symmetry, however from the worldvolume perspective of the NS$5$-branes, it still possesses the $SO(5,1)$ Lorentz invariance. Another characteristics of little string theory is the existence of Hagedorn density of states at temperature $T_{h}$, known as the Hagedorn temperature. Here the inverse of the Hagedorn temperature $\beta_{h}$ serves as the nonlocal scale of the theory \cite{Chakraborty:2020fpt}. The nonlocal effects are manifest at length scale less than the inverse Hagedorn temperature $\beta_{h}$ while above $\beta_{h}$ one recovers the usual behavior as observed for the local theory.

Nonlocal quantum theories often shows interesting behavior for physically measurable quantities at certain length scales. As an example, the holographic computation of entanglement entropy (using the Ryu-Takayanagi prescription \cite{Ryu:2006bv}) of certain nonlocal theories with appropriate gravity dual shows volume law behavior for the leading UV diverging term instead of the usual area law below some critical length scale of the theory \cite{Barbon:2008ut, Karczmarek:2013xxa, MohammadiMozaffar:2017nri}. A similar conclusion was also made by the authors in \cite{Fischler:2013gsa} regarding the computations of entanglement entropy and holographic mutual information in large $N$ strongly coupled noncommutative gauge theory which possesses an inherent non locality due to the natural UV/IR mixing that the theory exhibits. Regarding the scrambling of quantum information it is observed that the nonlocal theories can scramble information much faster compare to the local theories such that the thermalisation occurs at smaller time scale in theories with nonlocal interactions \cite{Lashkari:2013iga, Edalati:2012jj}. The same conclusion was also made in \cite{Fischler:2018kwt} regarding the information scrambling in noncommutative geometry by explicitly computing the butterfly velocity. It was shown that the butterfly velocity keeps on increasing with the noncommutative parameter. In this paper we also observe the increasing behavior of butterfly velocity at length scale less than the inverse Hagedorn temperature. However unlike the case in noncommutative geometry, in this case the butterfly velocity does not violate the bound imposed by the speed of light. This is because although the theory of little string is nonlocal, it still possess the lorentz symmetry as already discussed.

So far in our discussion we talked about the diagnosis of quantum chaos from the study of out of time ordered correlator. However calculation of OTOC is sometimes hard, especially when one tries to evaluate at finite temperature. It is recently observed that the chaotic nature of a many body quantum system has a direct manifestation in the energy density retarded Green's function \cite{Blake:2017ris, Blake:2018leo}. In other words the exponential growth of the OTOC (\ref{OTOC2}) can be realized from the structure two point correlation function involving the temporal component of the stress energy tensor on the boundary. According to this phenomenon which is known by the name `pole skipping', quasinormal frequency ($\omega$) and momentum ($k$) for collective excitations upon analytic continuation can take special complex values such that the residue of retarded Green's function in momentum space vanishes exactly. That is, at those special values of $\omega$ and $k$, line of poles and line of zeroes of retarded Green's function passes through each other and hence it is not uniquely defined. This phenomenon was first observed in a numerical computations for a particular holographic system \cite{Grozdanov:2017ajz} and then derived in \cite{Blake:2017ris} for an effective field theory of quantum chaos. The gravitational origin of this phenomenon was nicely explained in \cite{Blake:2018leo}. It was observed that at special complex values of $\omega$ and $k$, there exists two linearly independent ingoing solution for any bulk field equation of motion near the horizon so that one can define any arbitrary linear combinations of them to construct another possible solution. Hence the retarded Green's function which is determined by the unique choice of the ingoing solution, is not well defined due to the arbitrariness of the solution at the horizon. A series of computations has been done recently regarding the explicit computation of pole skipping points for different theories including the holographic ones.    \cite{Blake:2019otz,Grozdanov:2019uhi,Natsuume:2019xcy,Ceplak:2019ymw,Natsuume:2019vcv,Wu:2019esr,Ahn:2020bks,Grozdanov:2020koi,Ahn:2020baf,
Kim:2020url,Natsuume:2020snz,Abbasi:2020xli,Ceplak:2021efc,Kim:2021hqy,Choi:2020tdj,Haehl:2018izb,Jensen:2019cmr,Das:2019tga,Haehl:2019eae,
Ramirez:2020qer,Grozdanov:2018kkt,Natsuume:2019sfp,Li:2019bgc,Abbasi:2019rhy,Abbasi:2020ykq,Jansen:2020hfd,Sil:2020jhr,Yuan:2020fvv,
Jeong:2021zhz,Ahn:2019rnq,Blake:2021hjj,Liu:2020yaf}. In this paper we intend to carry on similar calculation for the holographic dual description of little string theory. We will be particularly interested in the pole skipping points occurring on upper half of the complex plane which are directly related to the parameters of chaos.

The rest of the paper is organized as follows, in section-{\bf 2} we briefly discuss the holographic dual of little string theory and the corresponding nonlocal scale. In section-{\bf 3} and {\bf 4}, we compute the holographic entanglement entropy and the entanglement wedge cross section to show directly the evidence of nonlocality at length scale smaller than some critical one. Section-{\bf 5} is entirely based on the computation of butterfly velocity using the gravitational shockwave analysis in a typical entangled state of two copies of the theory, known as the thermofield double state. In section-{\bf 6}, we study the disruption of left-right entanglement in the same thermofield double set up by computing the two sided holographic mutual information. In section-{\bf 7}, we evaluate the parameters of quantum chaos, namely the Lyapunov exponent and the butterfly velocity using the phenomenon of pole skipping. Finally, we conclude in section-{\bf 8}.

\section{Holographic description of little string theory}
There exists two different classes of non-gravitational theories in higher dimension that results in some limit of type-II string theory. The usual local field theories are one such class that arises in the low energy limit defined by $E\ll m_{s}$, $m_{s}=\frac{1}{\sqrt{\alpha^{\prime}}}$ being the string scale. On the other hand, there are consistent theories obtained in the limit of vanishing string coupling, $g_{s}\rightarrow 0$, but at finite string scale $m_{s}$. This constant string scale is important for the dynamics of these theories, known as the `little string theory'. It is a nonlocal theory defined on the worldvolume of $N\gg 1$ parallel NS$5$ branes after one considers the decoupling limit as mentioned above. In the vanishing string coupling limit however the coupling of the field theory on the NS$5$ branes remains finite and can be realized by considering the low energy limit \cite{Kutasov}. In string frame the supergravity solution can be described in terms of the following ten dimensional metric, the dilaton and a three form flux corresponding to the NS $B$ field \cite{Horowitz:1991cd,Maldacena:1997cg,Parnachev:2005hh,Chen:2013nma}
\begin{equation}\label{metricsf}
\begin{split}
ds_S^2&=-f(r)dt^2+dx_5^2+A(r)\Big(\frac{dr^2}{f(r)}+r^2d\Omega_3^2\Big),~~\text{where}\\
A(r)&=1+\frac{N\alpha^\prime}{r^2},~~f(r)=1-\frac{r_{H}^2}{r^2},~~e^{2\Phi}=gs^2A(r),\\
H_3&=2L\sqrt{N\alpha^{\prime}(r_H^2+N\alpha^{\prime})}V_3
\end{split}
\end{equation}
where $V_3$ is volume of the compact three sphere and $r_{H}$ is the black hole horizon. At energy scale $E\sim m_{s}$, the string scale, little string theory has a density of state which is larger compared to that of the standard local QFT known as the Hagedorn density of states. It is proportional to $e^{\beta_{h}E}$, $E$ being the energy density and $\beta_{h}$ as the inverse Hagedorn temperature.
Further, we define $L=\sqrt{N\alpha^\prime}$ such that the inverse Hagedorn temperature is given as,
\begin{equation}
\beta_{h}=2\pi \sqrt{N\alpha^{\prime}}=2\pi L.
\end{equation}
This inverse Hagedorn temperature scale determines the nonlocality of little string theory, such that at any scale lower than that the nonlocal phenomenon can be explicitly observed. In the rest of the paper we consider large value of $L$ (as $\beta_{h}$ is proportional to $L$) to show the nonlocal effects on different measures of information spread and also quantum chaos. In other words at separation smaller than $\beta_{h}$, there exist no local operators but they actually smeared over finite distance scale which is at least $\sim \beta_{h}$. The most interesting thing about the little string theory is that it possesses several characteristics of a critical string theory but is devoid of the nontrivial features of gravity. Using the above dual gravitational description we will highlight some of the interesting properties of little string theory relevant to the ongoing research in quantum information theory.
\section{Holographic Entanglement Entropy}
\begin{figure}[h]
\centering
\begin{tabular}{c}
\includegraphics[width=.55\textwidth]{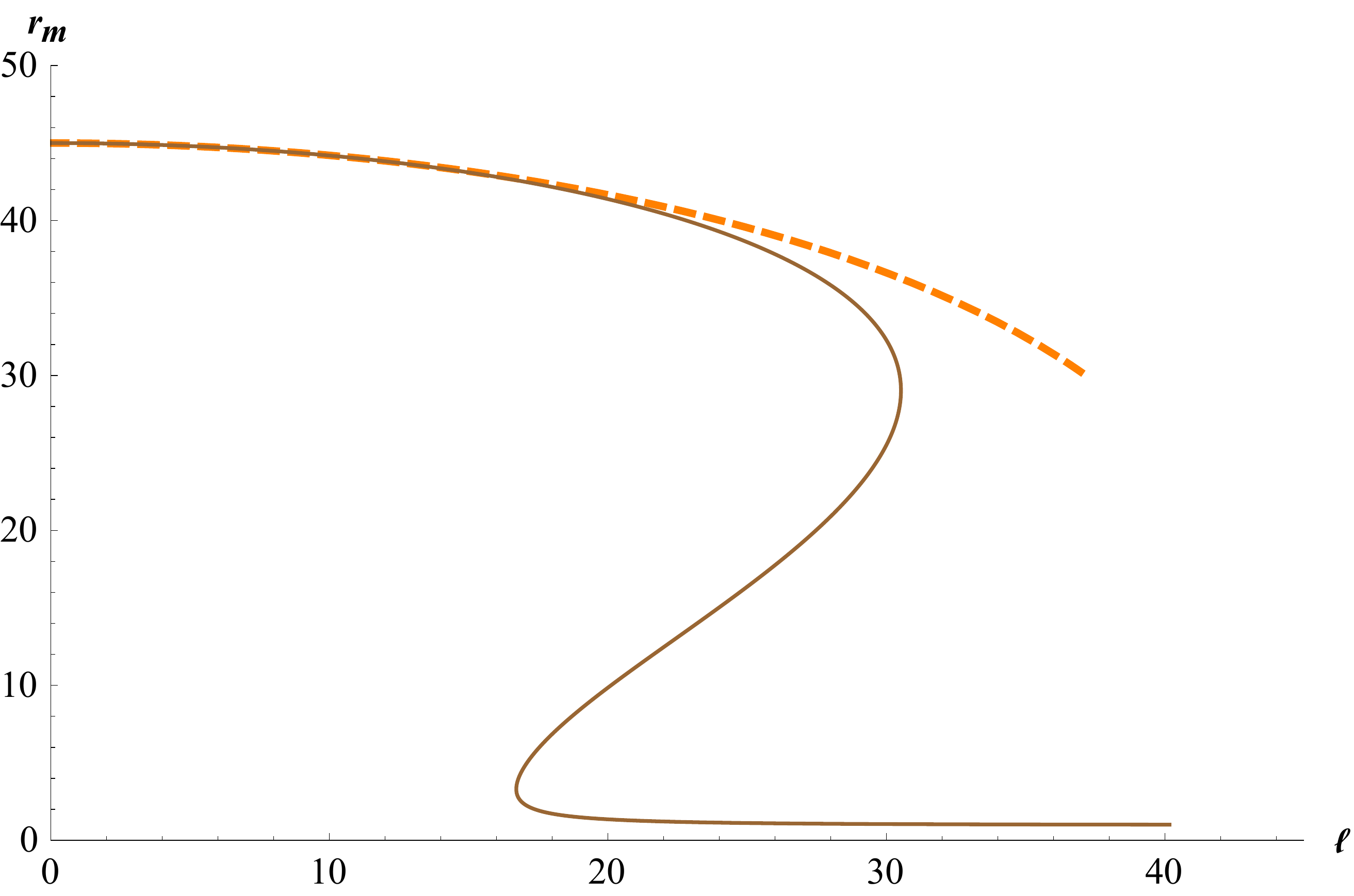}
\end{tabular}
\caption{Figure showing the penetration of the extremal surface in the bulk as a function of the strip width. The numerical result is denoted by the solid brown curve while the analytic result is indicated by the orange dashed line. Note that the analytic result matches nicely with the numerical one for lower values of the strip width $l$. In this plot we have fixed numerical value the ratio $r_{H}/L$ to $0.1$.}
\label{rm1}
\end{figure}

The Bekenstein-Hawking formula for the entropy of a black hole \cite{Bekenstein:1973ur} relates the entropy to some geometry of the black hole geometry, it scales as the area of the black hole horizon. The holographic prescription by Ryu-Takayanagi \cite{Ryu:2006bv} is a kind of generalisation of the Bekenstein-Hawking formula which states that the entropy of any arbitrary subsystem of a quantum system also has a geometrical interpretation, it equals the area of some minimal hypersurface embedded into the dual spacetime background. For a simple bipartite system, the entanglement entropy measures the strength of correlation between the degrees of freedom of any subsystem $\mathcal{A}$ with the rest of the system. Given a pure state $|\psi>$ of the whole system, the entanglement entropy of any subsystem $\mathcal{A}$ is defined as the Von-Neumann entropy,
\begin{equation}
S(\mathcal{A})=-Tr\left(\rho_{\mathcal{A}}\log{\rho_{\mathcal{A}}}\right),
\end{equation}
where, $\rho_{\mathcal{A}}$ is the reduced density matrix for $\mathcal{A}$.
In this section we will use the above mentioned Ryu-Takayanagi conjecture to compute the entanglement entropy of an infinitely long strip like entangling surface extending along the $(x_2,x_3,x_4,x_5)$ directions but with finite width $l$ along $x_1$ such that, $l/2\le x_{1}\le l/2$, $-R/2\le x_{2},...,x_{5}\le R/2$ with $R\rightarrow \infty$. Due to the varying dilaton profile in this case we will use the generalized version of the formula for the entanglement entropy of a subregion $A$ as given in \cite{Ryu:2006ef},
\begin{equation}\label{EE}
S(A)=\frac{\textrm{Area}(A)}{4G^{10}_N}=\frac{1}{4G^{10}_N}\int dx^{8} e^{-2\phi}\sqrt{g^{(8)}_{ind}},
\end{equation}
where, $G^{10}_N=8\pi^6\alpha^{\prime 4}$ is the $10$-dimensional Newton's constant and $g_{ind}$ is the metric induced on the strip which is embedded into the spacetime geometry following the equation $x_1=x_1(r)$ with time $t$ remaining constant throughout. Using the string frame metric (\ref{metricsf}), one can extremize the area functional in (\ref{EE}) with respect to the embedding function $x_1(r)$ to get the equation for the hypersurface extended in the bulk as,
\begin{equation}\label{dx1r}
\frac{dx_1}{dr}=\sqrt{\frac{A(r)}{f(r)}}\frac{1}{\sqrt{\frac{r^6 A(r)}{r_{m}^6 A(r_{m})}-1}},
\end{equation}
where $r_{m}$ denotes the turning point for the extremal surface, $\left(dx_1/dr\right)|_{r_{m}}\rightarrow \infty$. We will also define a UV cut-off at $r=r_{\delta}$ to regularize the area functional defined as, $x_{1}(r_{\delta})=\pm l/2$ which leads to the following expression for the strip width $l$ as,
\begin{equation}\label{l}
l=2\int_{r_{m}}^{r_{\delta}}dr~\sqrt{\frac{A(r)}{f(r)}}\frac{1}{\sqrt{\frac{r^6 A(r)}{r_{m}^6 A(r_{m})}-1}}.
\end{equation}
The final expression for the area of the subregion $A$ is given as,
\begin{equation}\label{A}
\mathcal{A}=\frac{2\Omega_{3}R^4}{g_{s}^2}\int_{r_{m}}^{r_{\delta}}dr~\frac{A(r)}{\sqrt{f(r)}}\frac{r^3}{\sqrt{1-\frac{r_{m}^6 A(r_{m})}{r^6 A(r)}}}.
\end{equation}
\begin{figure}[h]
\centering
\begin{tabular}{c}
\includegraphics[width=.55\textwidth]{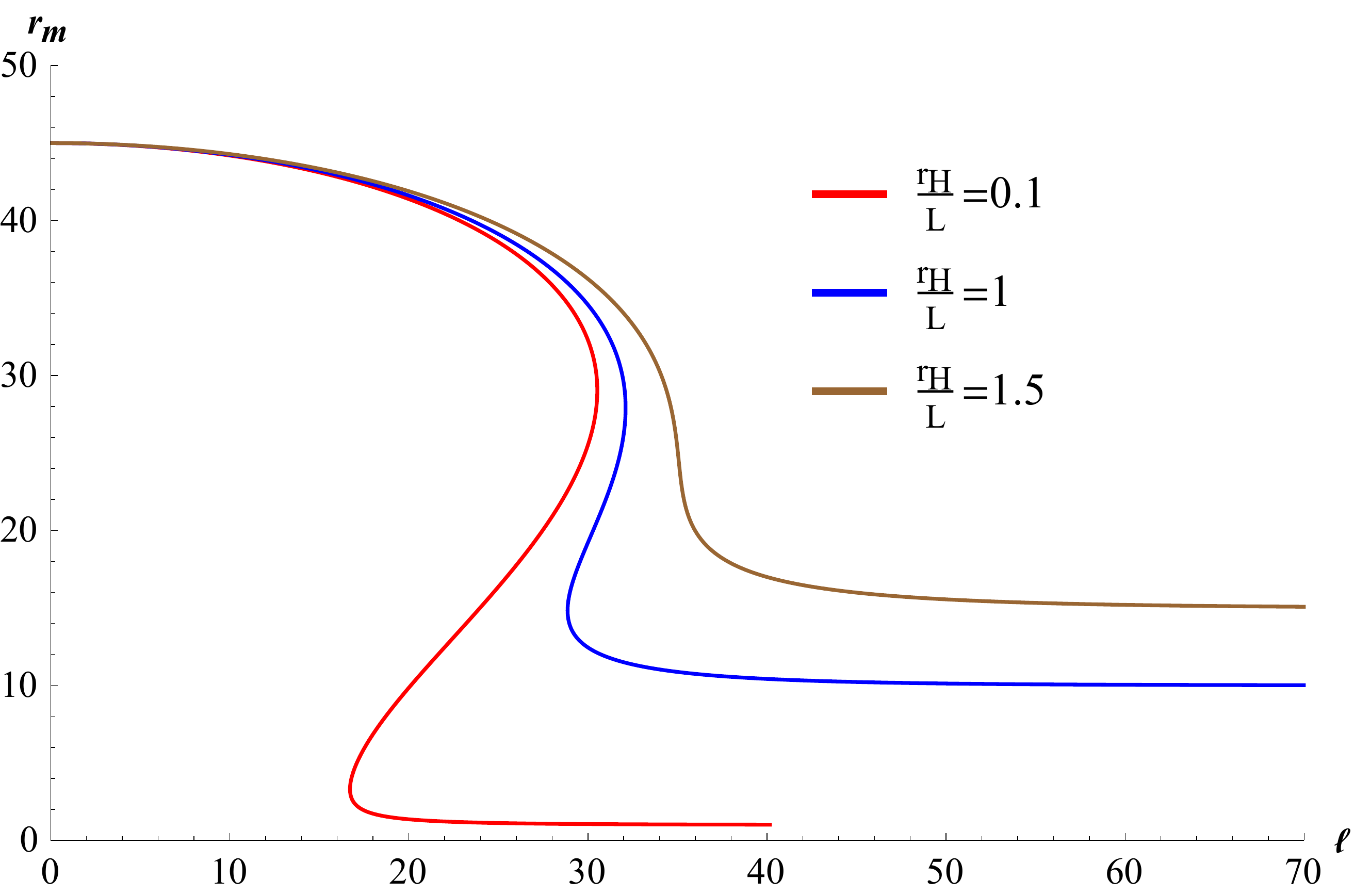}
\end{tabular}
\caption{Figure showing the embedding of the hypersurface as a function the strip width $l$ for different values of the horizon radius $r_{H}$ with fixed $L$. Here we have fixed $r_{\delta}$=45.}
\label{rm}
\end{figure}
Before going into the explicit computation of $S(\mathcal{A})$, let us first present a detailed discussion about the area functional as follows from the embedding of the extremal bulk hypersurface as a function of the strip width. As mentioned before, for a theory which exhibit nonlocal effects, there exists a critical value for the strip width, $l_{c}$ such that the usual area law behavior for the entanglement entropy is not obeyed and instead it follows the volume law when the width is less than $l_c$. In the following we will calculate an expression for this critical width which turns out to be proportional to the UV cut-off scale $r_{\delta}$. So if we set $r_{\delta}$ to infinity then for every finite width of the strip we will find a volume law behavior. However when we keep the UV cut-off scale at large but finite value, a phase transition regarding the behavior of entanglement entropy can be observed as the width of the strip is increases from a value less than $l_c$ to the one greater than the same \cite{Klebanov:2007ws}. In particular, the entanglement entropy will show a transition from the volume law to the usual area law behavior.
\paragraph{Behavior at UV scale:} Let us consider the plot as given in figure-\textbf{\ref{rm}} . We observe that at small values of $l$, that is for very narrow strip, there exist only one smooth extremal hypersurface which lies very close to the UV cut-off scale $r_{\delta}$. Also, in this smaller width region all the three hypersurface corresponding to different values of $r_{H}/L$ has the same form as shown in figure-\textbf{\ref{rm}}. In order to get an analytic expression for the hypersurface at small width one can use equation (\ref{dx1r}) to obtain the following relation between $r$ and $x_1$,
\begin{equation}\label{sl}
r(x_1)-r_m=\frac{\left(r_{m}^2-r_{H}^2\right)\left(2L^2+3r_{m}^2\right)}{8r_{m}\left(L^2+r_{m}^2\right)^2}x_{1}^2+\mathcal{O}\left(x_{1}^4\right).
\end{equation}
Using the boundary condition, $x_{1}(r_{\delta})=\pm l/2$ in the above equation, one can plot the variation of the extremal surface with strip width for small values of $l$. In figure-\textbf{\ref{rm1}}, we have shown the relationship between $r_{m}$ and $l$ using both the numerical result and the analytic expression (\ref{sl}) as obtained for small strip width.
Substituting $r(x_1)$ from equation (\ref{sl}) into the area functional in equation (\ref{A}), we obtain the approximate analytic expression for the area at small $l$ region to be,
\begin{equation}\label{UVA}
\mathcal{A}=\frac{\Omega_{3}R^4}{g_{s}^2}\biggl(l r_{\delta}^3+\frac{L^2}{2}l r_{\delta}+\frac{15}{128}l^3 r_{\delta}+\mathcal{O}\left(l^5\right)
\biggr)
\end{equation}
In a general black hole background that corresponds to an uniform thermal state in the boundary, the most dominant contribution to the entanglement entropy (basically the divergent part) comes from the correlation between the degrees of freedoms which are lying very close to the boundary of the entangling surface. Hence the most dominant part of entanglement entropy follows an area law. However there are also finite contributions to the entanglement entropy that follows a volume law behavior arising due to the correlation between the degrees of freedom from the entire volume of the entangling region.
However in this case we notice that the leading order term in $r_{\delta}$ in the above expression for the area depends on the size $l$ of the system, so that even the diverging part of the entanglement entropy shows an extensive behavior (volume law) which can be linked to the nonlocal nature of the underlying theory.
\paragraph{Behavior at IR scale:} As it follows from figure-\textbf{\ref{rm}} that there exists three solutions for the extremal surface at a given strip width around the critical value $l=l_{c}$. Among these three surfaces, the one at the middle always shows increasing behavior and hence it is not physical. The surface at the bottom has minimum area compared to the surface at the top and according to the RT prescription the bottom surface will be favourable. At even larger strip width there exists unique solution for the extremal surface and it will stay close to the horizon and far away from the UV cutoff. As a result the area functional in this scale is independent of the position of the UV cutoff and shows the usual area law behavior for the leading divergent part of the entanglement entropy. The area of the extremal surface in this case is given as,
\begin{equation}\label{IRA}
\mathcal{A}=\frac{\Omega_{3}R^4}{g_{s}^2}\biggl(r_{\delta}^4+\frac{2L^2+r_{H}^2}{2}r_{\delta}^2
+\frac{1}{4}\left(4L^2r_{H}^2+3r_{H}^4\right)\log{r_{\delta}}+\mathcal{O}\left(1/r_{\delta}\right)\biggr).
\end{equation}
The above result for the area is independent of the width $l$ of the infinite strip. Also with this result the entanglement entropy now follows an area law behavior as observed for a local theory. However, keeping the width of the strip fixed if one takes the limit $r_{\delta}\rightarrow \infty$, then the area again increases extensively similar to equation (\ref{UVA}).

The transition from a volume law behavior to an area law for the entanglement entropy occurs continuously as one moves from UV to IR scale. Hence one can calculate the critical length scale $l_{c}$ by comparing the two expressions for the area at high and low energy scales as given in (\ref{UVA}) and (\ref{IRA}), respectively and then solving for $l$.
\begin{equation}
l_c=\frac{r_{\delta}}{2}+\frac{3L^2+2r_{H}^2}{4r_{\delta}}+\mathcal{O}(1/r_{\delta}^2).
\end{equation}
So we see that the critical length scale is proportional to the UV cutoff which is a signature of nonlocal behavior of the theory.
The contribution to the entanglement entropy coming from different energy scales differs nontrivially. Also in standard local theory, the UV contribution to the entanglement entropy do factorize from that coming from the IR and they just adds up to the total entanglement entropy. In particular, the UV part of the entanglement entropy depends on the UV-cutoff $r_{\delta}$ and also it is local in a sense that only the degrees of freedom very close to the boundary of the subsystem gets involved to the quantum correlation across the boundary which results in an area law behavior. Conversely, the IR contribution is not only independent of the UV cutoff but does not depend on the scheme of the regularization either such that a good cutoff will only change the physics of the UV without affecting the IR. However, in a nonlocal theory the above mentioned factorisation between contributions coming from UV and the IR scales is no longer a valid assumption. For instance, as we have found, below the critical length scale the UV contribution of the entanglement entropy depends on the width $l$ of the infinite strip which is the IR scale of our theory. Hence the UV part of the entanglement entropy is no longer local and for a sufficiently narrow strip the degrees of freedom living on the entire strip now correlates to the one outside and hence it develops a dependance on the IR scale. This dependance on $l$ also leads to the violation of the area law. Finally, the critical length $l_{c}$ for the strip width also depends on the UV cutoff. So if the cutoff is taken at infinity, then a volume law for the entanglement entropy would follow for all energy scales. All the above results indicates the UV/IR mixing of the underlying nonlocal theory.

\begin{figure}[h]
\centering
\begin{tabular}{c}
\includegraphics[width=.55\textwidth]{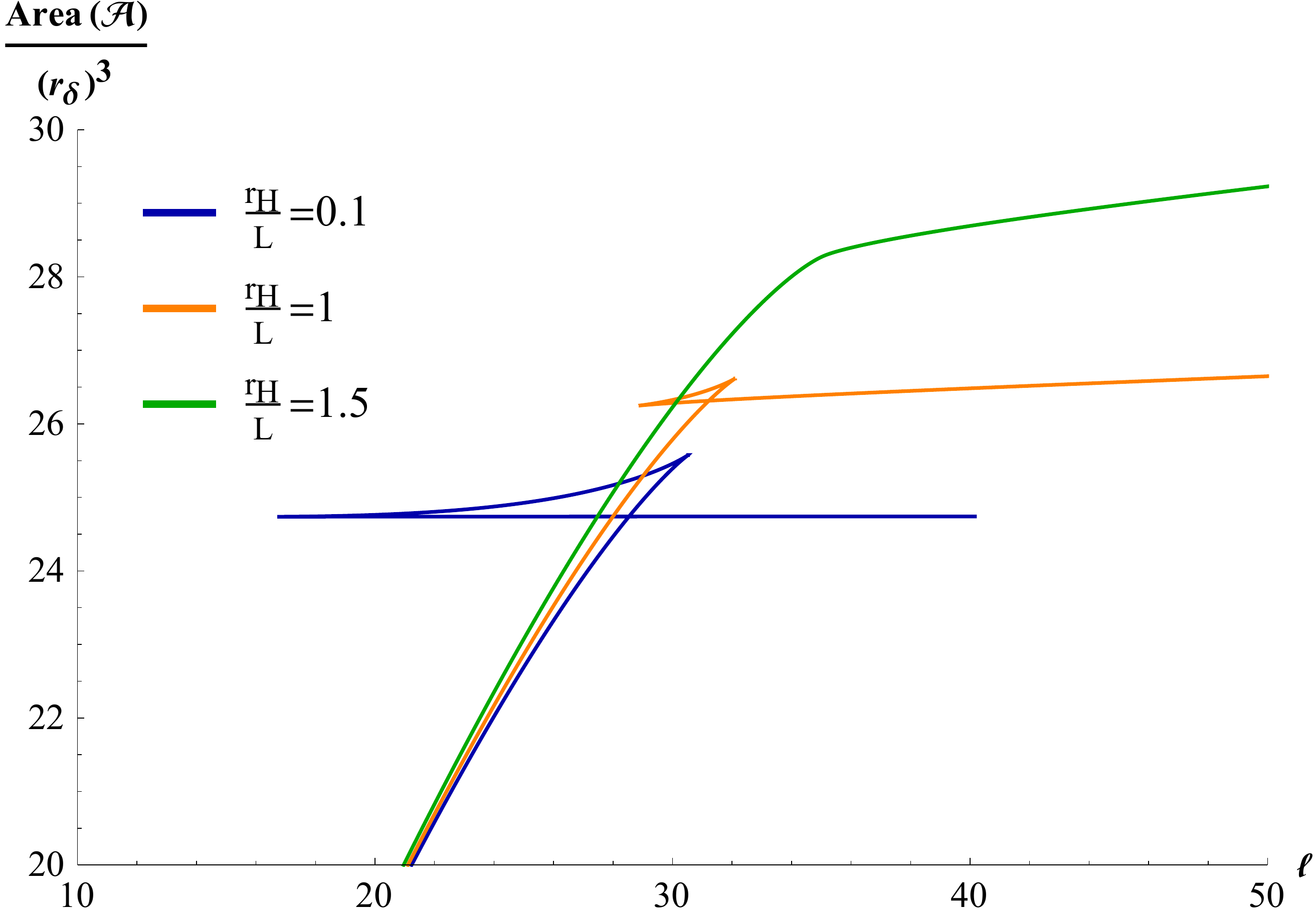}
\end{tabular}
\caption{Figure showing the variation of entanglement entropy with strip length $l$ for three different values of $r_{H}/L$.}
\label{Al}
\end{figure}

In figure-\textbf{\ref{Al}} we have shown the variation of the area with respect to the width $l$ varying from very small value to sufficiently large one. The area increases very quickly till the critical length scale and after that it slows down and eventually saturates. The transition from volume law to area law behavior is clearly realized from the same plot. Also note that at the critical value of strip width, the entanglement entropy is continuous but it's first derivative suffers a discontinuity indicating a first order phase transition. However as we increase the size of the black hole by considering larger value of $r_{H}$, there is no such phase transition which is explicitly shown in the same figure (see the green colored plot).
\section{Holographic entanglement wedge cross section}
In the last section we analyse the entropy of entanglement for a bipartite system in a pure state. However, for total correlation in a bipartite mixed state the appropriate measure as defined in quantum information theory is the entanglement of purification (EoP) \cite{terhal}. In the context of AdS/CFT correspondence, the geometrical quantity which is dual to EoP is conjectured in \cite{Takayanagi:2017knl, Nguyen:2017yqw} to be the Entanglement wedge cross section (EWCS). In holographic theories the computation of EWCS can be done in a straight forward manner by evaluating the area of the minimal hypersurface $\Gamma_{W}$ (see figure-{\bf }) inside the entanglement wedge in the bulk corresponding to the union of two subregions $\mathcal{A}$ and $\mathcal{B}$ on the boundary. In this section we adopt the holographic technique to calculate EWCS for the same gravitational background dual to the theory of little string theory. The goal is to see if the nonlocal nature of little string hteory has any direct manifestation in such measure of correlation in a mixed state.

Let us start with the technical definition of Eop.
Consider two subregions $\mathcal{A}$ and $\mathcal{B}$ in the form of a infinite strip of equal width $l$ on two separate the boundary slices at a constant time. In general, the two subregions will be in mixed states defined by the density matrix $\rho_{\mathcal{A}}\in \mathcal{H}_{\mathcal{A}}$ and $\rho_{\mathcal{B}}\in \mathcal{H}_{\mathcal{B}}$ respectively, where $\mathcal{H}_{\mathcal{A}}$ and $\mathcal{H}_{\mathcal{B}}$ are the corresponding Hilbert spaces. The total density matrix for the two bipartite systems $\rho_{\mathcal{A}\mathcal{B}}\in \mathcal{H}_{\mathcal{A}}\otimes\mathcal{H}_{\mathcal{B}}$ can be obtained from the pure state $|\psi>\in \mathcal{H}_{\mathcal{A}}\otimes\mathcal{H}_{\mathcal{B}}$ following the usual definition of density matrix, $\rho_{\mathcal{A}\mathcal{B}}=Tr_{\bar{\mathcal{A}}~\bar{\mathcal{B}}}|\psi><\psi|$. In the last definition $\bar{\mathcal{A}}$ and $\bar{\mathcal{B}}$ represents the additional degrees of freedoms required for the purification of $\rho_{\mathcal{A}\mathcal{B}}$ such that the state $|\psi>$ belongs to the extended Hilbert space $\mathcal{H}_{\mathcal{A}\bar{\mathcal{A}}}\otimes \mathcal{H}_{\mathcal{B}\bar{\mathcal{B}}}$. The EoP is defined as the minimum von-Neumann entropy $S(\mathcal{A}\bar{\mathcal{A}})$, minimized over all possible splitting of the purification into $\bar{\mathcal{A}}~\bar{\mathcal{B}}$,
\begin{equation}
E(\rho_{\mathcal{A}\mathcal{B}})=\mathop{min}_{\rho_{\mathcal{A}\mathcal{B}}=Tr_{\bar{\mathcal{A}}~\bar{\mathcal{B}}}|{\psi}\rangle \langle{\psi|}}S(\mathcal{A}\bar{\mathcal{A}}).
\end{equation}
\begin{figure}[h]
\centering
\begin{tabular}{c}
\includegraphics[width=.4\textwidth]{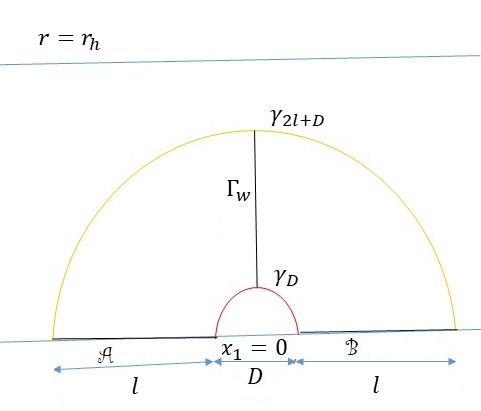}
\end{tabular}
\caption{A schematic representation of the holographic set up for the computation of EWCS.}
\label{EWD}
\end{figure}
In order to proceed with the holographic calculation of EWCS, we represent two subregions on the boundary as infinitely long strip with finite width $l$ in a similar fashion. The two subregions are separated by a distance $D$ as in figure-\textbf{\ref{EWD}}. The RT surface for $2l+D$ and $D$ are denoted by $\gamma_{2l+D}$ and $\gamma_{D}$ respectively. In a symmetric configuration around $x_1=0$, the EWCS is the minimal area of the hypersurface denoted by $\Gamma_{W}$ with the two end points coinciding on the RT surface $\gamma_{D}$ and $\gamma_{2l+D}$. The induced metric on the hypersurface is given as,
\begin{equation}
ds_{\Gamma_{W}}^2=\sum_{i=2}^{5}dx_{i}^2+A(r)\Big(\frac{dr^2}{f(r)}+r^2d\Omega_3^2\Big).
\end{equation}
The area of this minimal surface can be computed from the above induced metric yielding the final result for the EWCS as,
\begin{equation}\begin{split}
E_{W}&=\frac{\Omega_{3}R^4}{4 G_{N}^{(10)}g_{s}^2}\int_{r_{m}\left(2l+D\right)}^{r_{m}\left(D\right)}dr \frac{r^3 A(r)}{\sqrt{f(r)}},\\&
=\frac{\Omega_{3}R^4}{32 G_{N}^{(10)}g_{s}^2}\biggl[r\sqrt{r^2-r_{H}^2}\left(4L^2+2r^2+3r_{H}^2\right)
+r_{H}^2\left(4L^2+3r_{H}^2\right)\log{\left(r+\sqrt{r^2-r_{H}^2}\right)}\biggr]_{r_{m}\left(2l+D\right)}^{r_{m}\left(D\right)}
\end{split}
\end{equation}
Now, as we have seen in the previous section that the RT surface inside the bulk follow very different embedding depending on weather the width of the strip $l$ is greater or less than the critical length scale $l_{c}$ (see figure-\textbf{\ref{rm}}). So we expect the EWCS to show different behavior with increasing $l$ in regions where $l<l_{c}$ and $l>l_{c}$.
\begin{figure}[h]
\centering
\begin{tabular}{ccc}
\includegraphics[width=.45\textwidth]{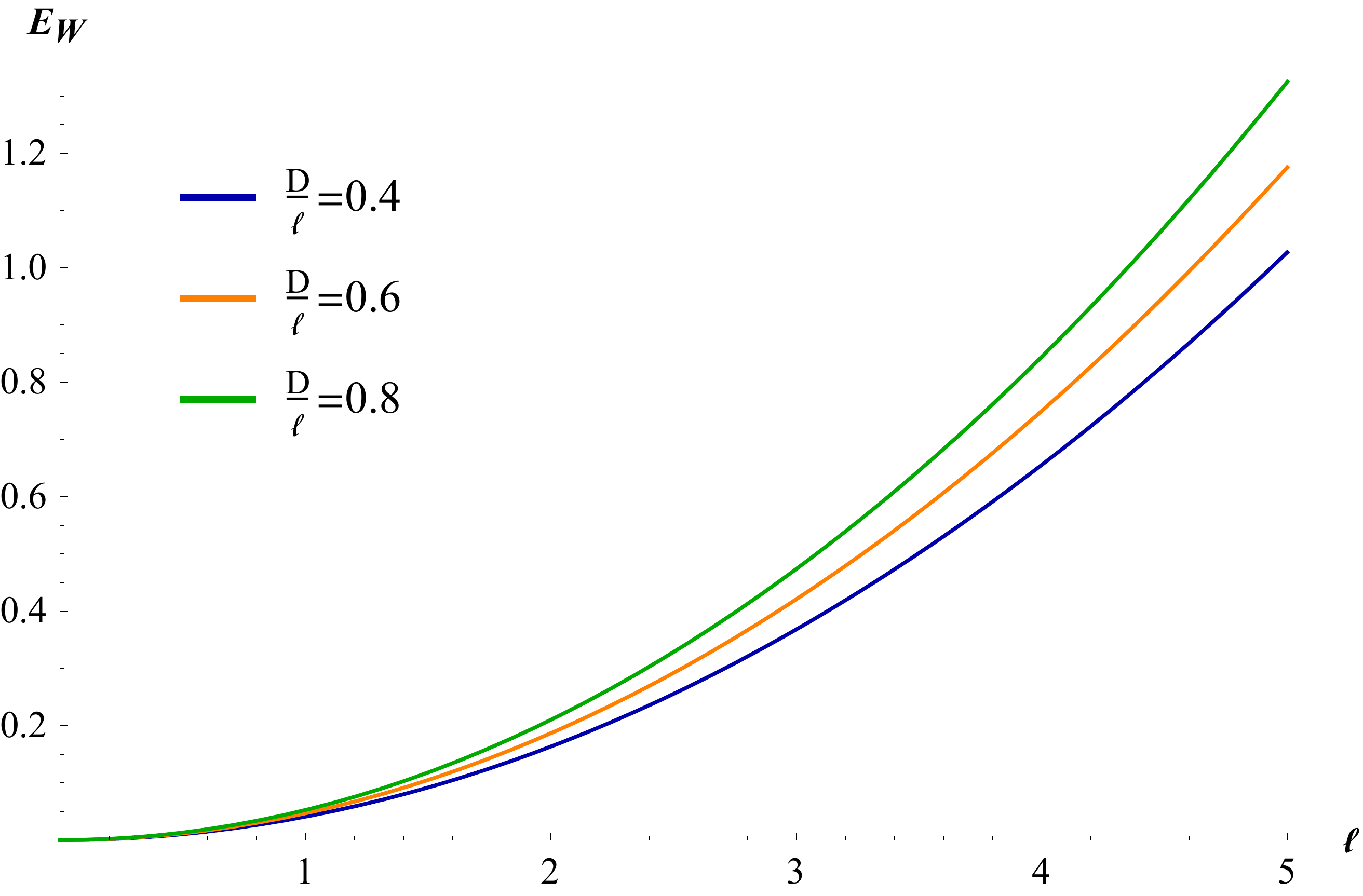}&&
\includegraphics[width=.45\textwidth]{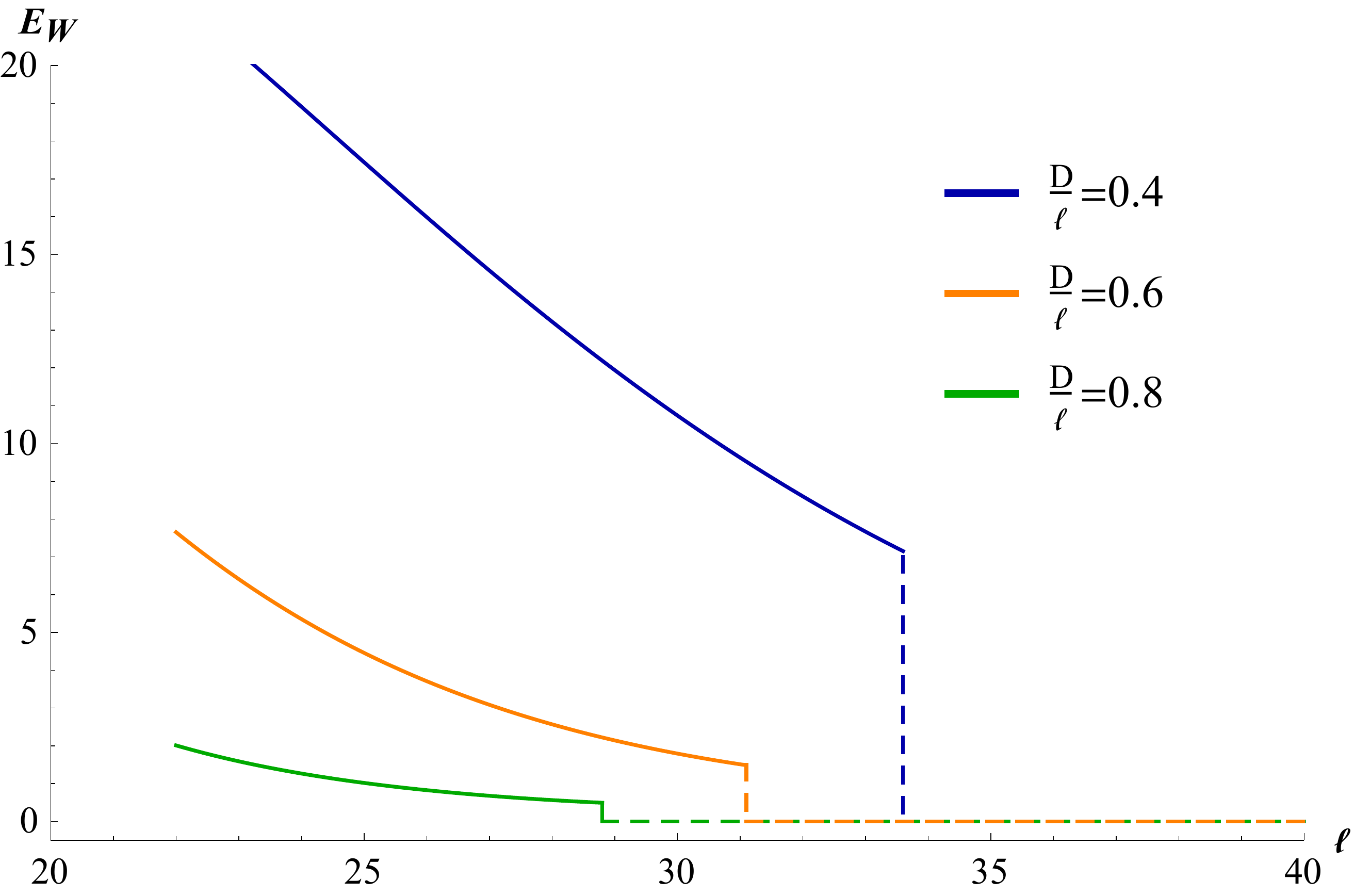}
\end{tabular}
\caption{a)Plot showing the variation of EWCS with strip width in the region of smaller values of $l<l_{c}$, b) Variation of EWCS with strip width in the region of larger $l>l_{c}$.}
\label{ew}
\end{figure}
In the left panel of figure-\textbf{\ref{ew}}, we see that the EWCS increases with the strip width $l$ for length scale which is less that the critical value $l_{c}$. This can be explain at least geometrically from the nature of the embedding of RT surface in the bulk geometry. Again referring to figure-\textbf{\ref{rm}}, we see that the RT surface for small values of $l<l_{c}$ stays very close to the boundary and as $l$ increases it starts to penetrate deep into the bulk towards the horizon. Now the RT surface $\gamma_{2l+D}$, corresponding to the region of total width $2l+D$ on the boundary, moves towards the horizon faster than the other RT surface $\gamma_{D}$ as the width $l$ is increased. This is clearly evident from the increasing slope of the minimal surface in figure-\textbf{\ref{rm}} in the region of small $l$. Hence the length of the extremal surface $\Gamma_{W}$ increases with $l$ initially. However, as the width is increased to sufficiently large value beyond $l_{c}$, the minimal surface $\gamma_{2l+D}$ now reaches enough close to the horizon and can no longer penetrate anymore. On the other hand, the surface $\gamma_{D}$ being comparatively smaller in area, can still increase in size and move towards the horizon. As a result the length of the extremal surface $\Gamma_{W}$ decreases as shown in the right panel of figure-\textbf{\ref{ew}}.

Clearly the increase in EWCS in the region of small $l$ is due to the nonlocal correlation between the two subregions $\mathcal{A}$ and $\mathcal{B}$. In other words, for small strip widths, the degrees of freedoms from the entire volume of the two subregions participate in the correlation due to the nonlocal nature of the interaction and hence we see an increase in the correlation with $l$. However beyond the critical length scale the nonlocal correlation ceased and the usual decreasing behavior is observed.
As the strip width is increased beyond a particular value of $l$, the two subregions gets in a disconnected phase such that the correlation between them drops discontinuously to zero. \footnote{Here we have considered the infinitely long parallel strip configurations of width $l$ to calculate the EWCS. In this case there exists two different phases, namely (i) the connected phase and (ii) the disconnected phase which basically indicates the degree of correlation between two subsystems. In connected phase the correlation is finite, on the other hand in the disconnected phase there is no correlation between the two subsystems. Now, in the connected phase the area of the minimum cross-section is finite till some particular value of the strip width $l$, say $l_{con}$. However as the width is increased beyond $l_{con}$, the entanglement wedge doesn't exist anymore and as a result the corresponding minimum cross-section vanishes. So the reason for the disconnected phase is the sudden non-existence of the entanglement wedge as the value of $l$ gets bigger than $l_{con}$. Also, as already mentioned, physically the occurrence of the disconnected phase indicates zero correlation between the two subregions.}
\section{Study of shockwave geometries}
\label{section:shockwave}

\subsection{Gravity setup for Unperturbed Geometry}
We will start by considering the following $10$-dimensional metric (\ref{metricsf}) corresponding to an extended two sided black brane geometry,
\begin{equation}\label{TwosidedBrane}
ds^2=-G_{tt}dt^2+G_{rr}dr^2+G_{ij}dx^idx^j+G_{lm}d\theta^ld\theta^m,
\end{equation}
Where, ${i,j=1,2,3,4,5}$ represents the five dimensional flat directions and ${l,m=1,2,3}$ corresponds to the compact three dimensional unit sphere $S^3$. The asymptotic boundary is located at $r=\infty$ and the horizon at $r=r_h$, $r$ being the radial coordinate. Different metric components of the above equation (\ref{TwosidedBrane}) can be expanded near the horizon and are given as,
\begin{equation}
G_{tt}=Z0(r-rh),~G_{rr}=\frac{Z1}{r-rh},~G_{ij}=G_{lm}=\text{Constant}.
\end{equation}
The inverse hawking temperature can be calculated from the above data as
\begin{equation}
\beta=4\pi\sqrt{Z1/Z0}=1/T_H.
\end{equation}
For the gravitational shockwave analysis of the above geometry it is convenient to work with the Kruskal coordinate system $(U,V)$ which is defined below as,
\begin{equation}
UV=e^{\frac{4\pi}{\beta} r_{*}},~~U/V=-e^{-\frac{4\pi}{\beta} t}.
\label{Kruskal}
\end{equation}
where $r_{*}$ is the tortoise coordinate given below as,
\begin{equation}
r_{*}=\int\sqrt{\frac{G_{tt}}{G_{rr}}}dr,
\end{equation}
In terms of the Kruskal coordinate the metric (\ref{TwosidedBrane}) can be rewritten as,
\begin{equation}
ds^2=2F(U,V)dUdV+G_{ij}(U,V)dx^idx^j+G_{lm}(U,V)d\theta^ld\theta^m,
\label{kmetric}
\end{equation}
where $F(U,V)=\frac{\beta^2}{8\pi}\frac{G_{tt}(UV)}{UV}$. Now in terms of the $(U,V)$ coordinate, the two asymptotic boundaries on the left and the right side (see figure-\textbf{\ref{TFD}}) are situated at $UV=-1$, while at the horizon we have $UV=0$. Considering the above metric, Einstein's field equation can be written as,
\begin{equation}
R_{\mu\nu}-\frac{1}{2}\partial_{\mu}\partial_{\nu}\phi-\frac{1}{4}\exp(-\phi)H_{\mu bc}H_{\nu}^{bc}+\frac{1}{6(d-2)}g_{\mu\nu}\exp(-\phi)H_3^2=8\pi G_N T_{0,\mu\nu} ^M,
\label{Unpert}
\end{equation}
where, $T_{0,\mu\nu} ^M$ is the energy momentum tensor corresponding to the matter part of the action computed using the unperturbed geometry. It is presumed to have the following structure,
\begin{eqnarray}
T_{0,\mu\nu} ^M dx^\mu dx^\nu=&&2T_{UV}(U,V)dUdV+T_{UU}(U,V)dU^2+T_{VV}(U,V)dV^2+T_{ij}(U,V)dx^{i}dx^{j}\nonumber\\
&&+T_{lm}(U,V)d\theta^ld\theta^m.
\end{eqnarray}
\subsection{Shockwave geometries}
The two sided black hole geometry discussed above is holographically dual to the thermofield double state as already defined in the introduction section.
In this section we will discuss the possible modifications of the metric in (\ref{kmetric}) as a result of a localised perturbation on one of the boundaries (say left) at early time. In the gravity description this early perturbation corresponds to an energy pulse that originates from the past horizon and propagates into the future singularity. Due to the strong gravitational attraction, the energy of this pulse increases exponentially with time. For sufficiently early perturbation on the boundary the energy pulse produces a shockwave in the bulk geometry such that the corresponding backreaction becomes non negligible. This perturbed geometry is referred to as the shockwave geometry. Analysing the interaction of this shockwave with the black hole horizon one can extract important quantities like butterfly velocity $v_B$ and Lyapunov exponent $\lambda_L$ etc. These quantities can shed light on the chaotic features of holographic theories.

The shockwave we consider here is localized at $U=0$ and propagates along the $V$ direction resulting an extension along $V$ and a compression along $U$. The energy momentum tensor that describe the whole effect is given as,
\begin{equation}
T_{UU}^{Shock}=E\delta(U)e^{\frac{2\pi t}{\beta}}e^{i k.x}.
\label{shock}
\end{equation}
Here we will work in the momentum space due to the absence of any well defined gauge invariant local operators in position space for the underlying nonlocality of little string theory \footnote {For detailed discussion on this see \cite{Fischler:2018kwt}.}. To incorporate the backreaction in our calculation we made the replacement $V\rightarrow V+\alpha$. Here $\alpha$ captures key information about the profile of the shockwave,
\begin{equation}
\alpha=\tilde{\alpha}(t,k)e^{i k.x}.
\label{alpha}
\end{equation}
$\tilde{\alpha}(t,k)$ can be obtained as solution of the Einstein's field equation. To ensure the fact that only the region with $U>0$ is modified due to the shockwave, we replace $V$ with $V+\Theta(U)\alpha$, $\Theta(U)$ being the Heaviside step function.
After doing some algebraic manipulation, the modified metric and energy momentum tensor can be expressed respectively as,
\begin{equation}
ds^2=2FdUdV+G_{ij}dx^idx^j+G_{lm}d\theta^ld\theta^m-2A\delta(U)dU^2,
\label{pertmet}
\end{equation}
\begin{eqnarray}
T^{M}&&=2\biggl(T_{UV}-\alpha T_{VV}\delta(U)\biggr)dUdV+T_{VV}dV^2+\biggl(T_{UU}+\alpha^2 T_{VV}\delta(U)^2-2\alpha~T_{UV}\delta(U)\biggr)dU^2\nonumber\\
&&+T_{ij}dx^idx^j+T_{lm}d\theta^ld\theta^m.
\label{perten}
\end{eqnarray}
Finally, for perturbed geometry we can write Einstein's equation of motion as,
\begin{equation}
R_{\mu\nu}-\frac{1}{2}\partial_{\mu}\partial_{\nu}\phi-\frac{1}{4}\exp(-\phi)H_{\mu bc}H_{\nu}^{bc}+\frac{1}{6(d-2)}g_{\mu\nu}\exp(-\phi)H_3^2=8\pi G^{10}_N (T_{\mu\nu} ^M+T_{\mu\nu}^{shock}).
\label{pert}
\end{equation}
with $T^{M}$ and $T^{shock}$ is given in equation (\ref{shock}) and (\ref{perten}).
Now, we need to solve the $\text{UU}$ component of equation (\ref{pert}). Also notice that the components of the energy momentum tensor and metric should obey the following conditions \cite{Sfetsos:1994xa, Jahnke:2017iwi}
\begin{equation}
F_{,V}=G_{ij,V}=T_{VV}^{M}=0~~\text{at~U}=0.
\label{cond}
\end{equation}
We first solve equation (\ref{Unpert}) for $(T_{UV},T_{VV},T_{UU})$ and then substitute these back into the $\text{UU}$ component of (\ref{pert}) to finally obtain the following result for $\tilde{\alpha}$ given as,
\begin{equation}
\tilde{\alpha}\left(-Fk_ik_j+\frac{1}{2}G_{ij,UV}\right)G^{ij}=8\pi G^{10}_{N}E e^{\frac{2\pi t}{\beta}}.
\end{equation}
In terms of coordinate $(t,r)$, the above equation takes the following form
\begin{equation}
\left(G^{ij}k_ik_j+\tilde{M}^2\right)\tilde{\alpha}=-e^{\frac{2\pi\left(t-t_{*}\right)}{\beta}},
\label{beta}
\end{equation}
where
\begin{equation} \tilde{M}^2=\frac{4\pi^2}{\beta^2}\Bigg(\frac{1}{G_{tt}^{\prime}(r_h)}\Big(5\Big(\frac{G_{11}^{\prime}(r_h)}{G_{11}(r_h)}\Big)
+3\Big(\frac{G_{\theta\theta}^{\prime}(r_h)}{G_{\theta\theta}(r_h)}\Big)\Big)\Bigg),
\end{equation} and
\begin{equation}
t_{*}=\frac{\beta}{2\pi}\log\left(\frac{F(r_h)}{8\pi G^{10}_N E}\right).
 \end{equation}
where, $t_{*}$ is known as the scrambling time which indicates the time scale for saturation of complexity or growth of operator size. In other words, it denotes the time scale for the perturbation to spread over all boundary space time.
For diagonal metric $G_{ij}$, we can estimate the parameter $\tilde{\alpha}$ as,
\begin{equation}\label{ii}
\tilde{\alpha}(t,k)\sim\frac{e^{\frac{2\pi(t-t_{*})}{\beta}}}{G^{ii}(r_h)k_i^2+\tilde{M}^2}.
\end{equation}
For homogeneous shock, all the components of $k$ are same and we set it to zero such that the shock parameter $\tilde{\alpha}$ can be expressed as,
\begin{equation}
 \tilde{\alpha}(t,k)\sim \text{Constant}\times e^{(\frac{2\pi(t-t_{*})}{\beta})}.
 \end{equation}
We can compare it with the double commutator in (\ref{OTOC2}) and obtain the Lyapunov exponent $\lambda_L$ as,
\begin{equation}
\lambda_{L}=\frac{2\pi}{\beta}.
\label{Lyapunov}
\end{equation}
Notice that equation (\ref{ii}) has a pole at a particular value of $k$ proportional to $\tilde{M}$. The expression for $k^2$ evaluated at the pole is given in terms of $\lambda_{L}$ and $v_{b}$ as,
\begin{equation}
k^2=-\frac{\lambda_L^2}{v_b^2}.
\end{equation}
Using the above formula together with equation (\ref{Lyapunov}) we can calculate butterfly velocity as,
\begin{equation}
v_{b}^2=\frac{\lambda_L^2}{\tilde{M}^2}=\frac{1}{G_{11}(r_h)}\Bigg(\frac{1}{G_{tt}^{\prime}(r_H)}\Big(5\frac{G_{11}^{\prime}(r_H)}
{G_{11}(r_h)}+3\frac{G_{\theta\theta}^{\prime}(r_h)}{G_{\theta\theta}(r_h)}\Big)\Bigg)^{-1}.
\label{butterfly}
\end{equation}
For the little string theory butterfly velocity can be obtained by substituting the exact metric components from equation (\ref{metricsf}) into (\ref{butterfly}) to obtain the following result,
\begin{equation}
v_{b}^2=\frac{L^2+r_{H}^2}{2L^2+3r_{H}^2}.
\label{butterfly1}
\end{equation}
\begin{figure}[h!]
 \centering
 \includegraphics[width=8cm,height=5cm]{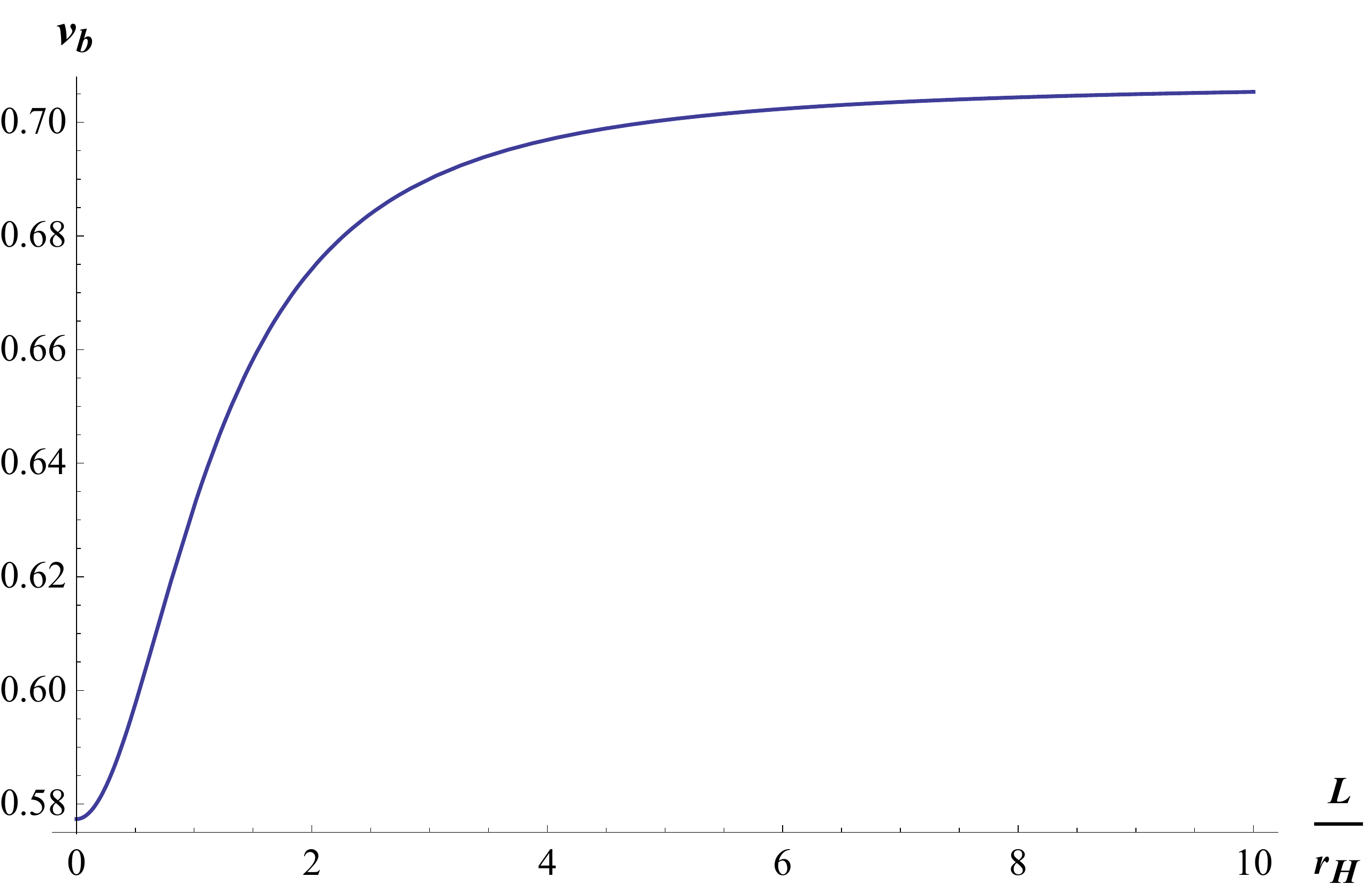}
 \caption{Variation of butterfly velocity $v_b$ with length scale $\frac{L}{r_H}$ }
 \label{butterflyvelocity}
 \end{figure}
In figure-\textbf{\ref{butterflyvelocity}}, we have plotted $v_b$ with respect to a dimensionless quantity $\frac{L}{r_H}$. It can be noticed from figure-\textbf{\ref{butterflyvelocity}}, that for increasing values of $\frac{L}{r_H}$, butterfly velocity $v_b$ increases sharply and finally saturates to a value less than the speed of light, $c=1$. The initial increase of $v_{b}$ with $\frac{L}{r_H}$ indicates a faster spread of information due to the nonlocal property of little string theory. The saturation of $v_{b}$ below the speed of light is due to the Lorentz invariance of the little string theory.
\section{Holographic mutual information of two sided geometry}
Mutual information plays an important role in defining nonlocal correlation between noninteracting and non-overlapping subregions. For two sided black hole geometry, these two subregions are considered at two asymptotic boundaries of the corresponding Kruskal diagram. The correlation between these two distant subregions can be realized through the inverse of the geodesic distance through the wormhole geometry extending from the left boundary to the right one. Due to the butterfly effect in holographic theories, a small perturbation in the asymptotic boundary can destroy the left-right entangled pattern completely as mentioned above. We will be interested in the process of this disruption, more precisely the rate with which it happens.


\subsection{Two sided mutual information for unperturbed geometry}
\begin{figure}[h!]
\centering
\begin{tabular}{c}
\includegraphics[width=.5\textwidth]{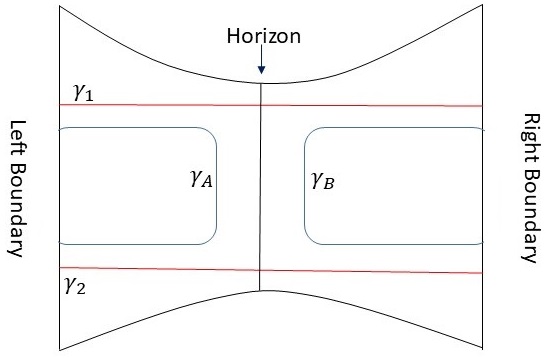}
\end{tabular}
\caption{Schematic diagram of unperturbed two sided geometry}
\label{untfd}
\end{figure}

\begin{figure}[h!]
  \centering
  \includegraphics[width=8cm,height=6cm]{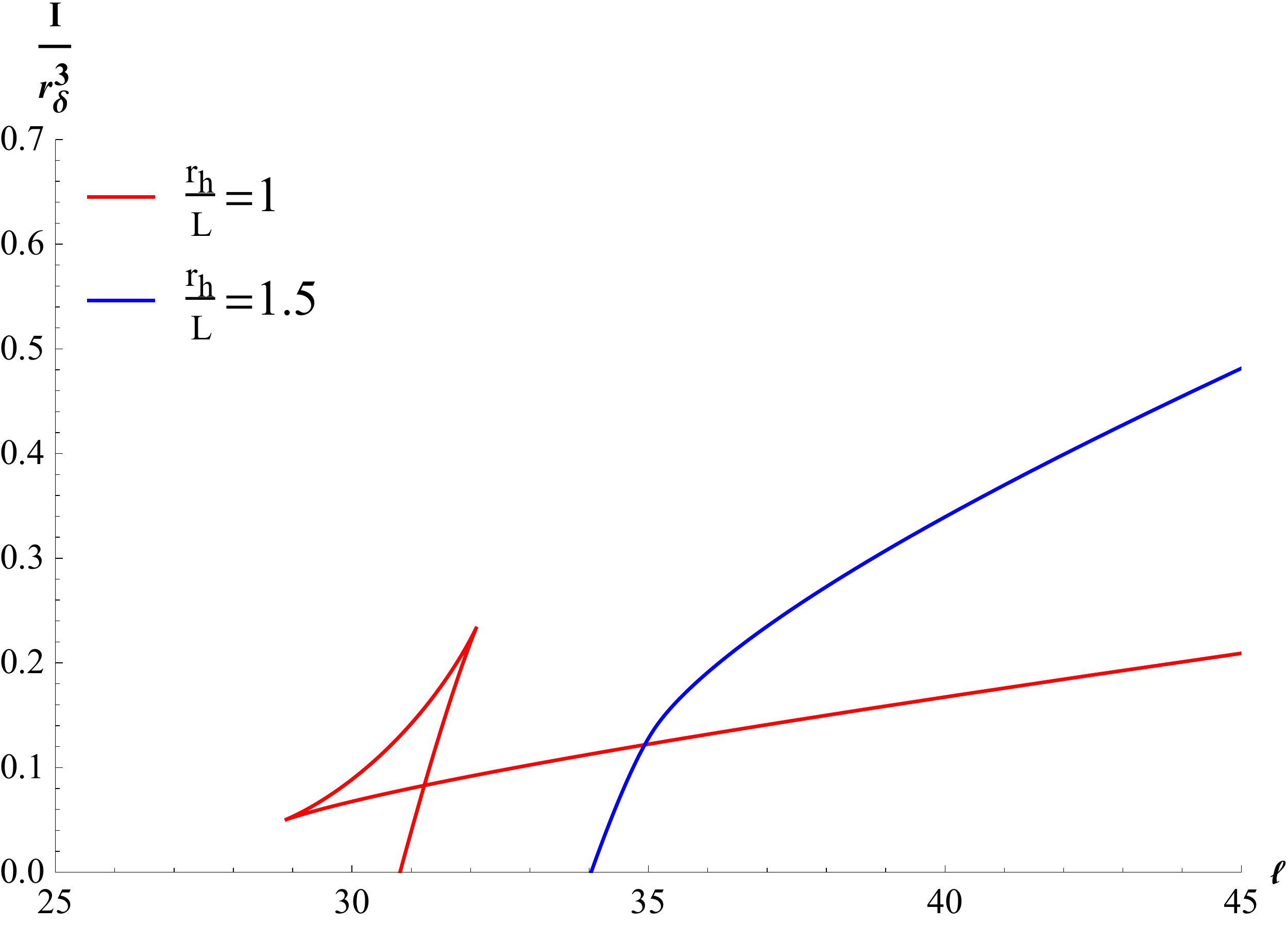}
  \caption{Plot showing variation of mutual information $\text{I}$ with strip width $l$}
  \label{Mutualinformation}
  \end{figure}

We will start with the computation of mutual information in the unperturbed gravity dual of little string theory using the HRRT prescription of \cite{Hubeny:2007xt}.
We consider two identical infinitely long strip like entangling surfaces $A$ and $B$ in the left and right boundaries extended along the spacial directions $(x_2,x_3,x_4,x_5)$ such that $-R/2\le x_{2,3,4,5}\le R/2$ with $R\rightarrow\infty$. However, it has a finite width $l$ along the $x_1$ direction. Considering a constant time slice, the corresponding Ryu-Takayanagi surfaces in the bulk exterior on both sides of the geometry is defined by the embedding $x_{1}(r)$. This embedding will generate a surface $\gamma_A$ in the left exterior and $\gamma_B$ in the right exterior. Also the corresponding bulk hypersurface for $A\cup B$ is given as $\gamma_{1}\cup \gamma_{2}$ as shown schematically by the red colored lines in figure-\textbf{\ref{untfd}}
\\
With this the mutual information $I(A,B)$ is defined as,
\begin{equation}
I(A,B)=S(A)+S(B)-S(A\cup B).
\label{MIU}
\end{equation}
where $S(A)$,~$S(B)$,~$S(A\cup B)$ are the entanglement entropy of $A$, $B$ and $A\cup B$ respectively. Using the HRRT formalism they can be calculated from the minimum area of the hypersurface $\gamma_A$,~$\gamma_B$,and $\gamma_1\cup\gamma_2$ respectively. The induced metric can be written as,
\begin{eqnarray}
g_{rr}&=&\left(G_{rr}+G_{11} x^{\prime}(r)^2\right),\nonumber\\
g_{\theta_{i}\theta_{i}}&=&G_{\theta_{i}\theta_{i}}\times \text{metric of}~ S^3,~~\text{i}=1,2,3,~~G_{11}=G_{22}=G_{33}=G_{44}=G_{55}.
\end{eqnarray}
The area functional for $\gamma_{A}$ can be calculated as,
\begin{eqnarray}
\mathcal{A}(\gamma_{A})&=&\int d^8\sigma\sqrt{(det g_{ab})},\nonumber\\
&=&\Omega_{3} R^4\int dr~ r^3 A^{5/8}\Big(\frac{A^{3/4}}{f}+\frac{x^{\prime}(r)^2}{A^{1/4}}\Big)^{1/2},\nonumber\\
&=&\Omega_{3} R^4\int dr \mathcal{F}(x,x^{\prime},r),
\label{area}
\end{eqnarray}
where $\Omega_{3}$ is the volume of three sphere and $R^4=\int dx_2dx_3dx_4dx_5$. The integrand in the last line of the above equation, $\mathcal{F}$, do not depend on $x$ explicitly hence the corresponding canonical momentum is a conserved quantity which can be calculated as,
\begin{equation}
P=\frac{\partial \mathcal{F}}{\partial x^{\prime}}=\frac{r^2 A(r)^{3/8}x^{\prime}}{\sqrt{\frac{A(r)^{3/4}}{f(r)}+\frac{{x^{\prime}}^2}{A(r)^{1/4}}}}.
\label{conserv}
\end{equation}
At turning point, $r=r_m$, $x^{\prime}$ will become infinitely large. Upon imposing this condition the conserved quantity $P$ is given as,
\begin{equation}
P|_{r=r_m}=r_m^3\sqrt{A(r_m)}.
\label{tur}
\end{equation}
Solving equation (\ref{conserv}) for $x^{\prime}$ along with the condition as given in (\ref{tur}) we obtain,
\begin{equation}
x^{\prime}=\sqrt{\frac{A(r)}{f(r)}}\frac{1}{\frac{r^6A(r)}{r_m^6A(r_m)}-1},
\label{prime}
\end{equation}
Substituting this expression of $x^{\prime}$ back into equation (\ref{area}) results in the following on shell area,
\begin{equation}
\mathcal{A}(\gamma_{A})=2\Omega_3R^4\int_{r_m}^{r_{\delta}}\frac{r^3A(r)}{\sqrt{f(r)}}\frac{1}{\sqrt{1-\frac{r_m^6A(r_m)}{r^6A(r)}}}dr,
\end{equation}
where $r_\delta$ is the UV cut-off as already defined. Finally, we can write down the corresponding entanglement entropy $S_A$ as,
\begin{equation}
S(A)=\frac{\mathcal{A}(\gamma_A)}{4G^{10}_N}=\frac{\Omega_3R^4}{2G^{10}_N}\int_{r_m}^{r_\delta}\frac{r^3A(r)}{\sqrt{f(r)}}\frac{1}{\sqrt{1-\frac{r_m^6A(r_m)}{r^6A(r)}}}dr.
\end{equation}
An exact similarly expression follows for the entanglement entropy of strip $B$. In order to obtain $S(A\cup B)$, we need to calculate the combined area of the surfaces passing through the interior connecting both sides of the geometry. There are two such surfaces, one at $x=0$ which is denoted as $\gamma_1$ and the other one is at $x=l$ denoted as $\gamma_2$. Using the symmetric configuration of these surfaces one can express the area of surface $\gamma_1\cup\gamma_2$ as,
\begin{equation}
\mathcal{A}(\gamma_1\cup\gamma_2)=4\Omega_3R^4\int_{r_h}^{r_\delta}\frac{r^3A(r)}{\sqrt{f(r)}}dr.
\end{equation}
Corresponding entanglement entropy can be written as,
\begin{equation}
S(A\cup B)=\frac{\Omega_3R^4}{G^{10}_N}\int_{r_h}^{r_\delta}\frac{r^3A(r)}{\sqrt{f(r)}}dr.
\end{equation}
Using these results in (\ref{MIU}) the final expression of mutual information can be written as the following integral form,
\begin{equation}
I(A,B)=\frac{\Omega_3R^4}{G^{10}_N}\Bigg(\int_{r_m}^{r_{\delta}}\frac{r^3A(r)}{\sqrt{f(r)}}\frac{1}{\sqrt{1-\frac{r_m^6A(r_m)}{r^6A(r)}}}dr-\int_{r_h}^{r_{\delta}}\frac{r^3A(r)}{\sqrt{f(r)}}dr\Bigg).
\label{MIL}
\end{equation}
On the other hand the strip width $l$ can be expressed as a function of $r_m$ and is given as,
\begin{equation}
l=\int dx=\int x^{\prime}dr=2\int_{r_m}^{r_{\delta}}\Bigg(\sqrt{\frac{A(r)}{f(r)}}\frac{1}{\frac{r^6A(r)}{r_m^6A(r_m)}-1}\Bigg)dr.
\end{equation}
In figure-\textbf{\ref{Mutualinformation}}, we have shown the variation of the mutual information $\text{I}$(A,B) with respect to the strip width $l$ for two different values of $\frac{r_h}{L}$. It is qualitatively very much similar to the variation of entanglement entropy with $l$ as discussed in section-{\bf 3} which is not surprising at all.
 \subsection{Two sided Mutual Information in presence of shock wave}
\begin{figure}[h!]
\centering
\begin{tabular}{c}
\includegraphics[width=.5\textwidth]{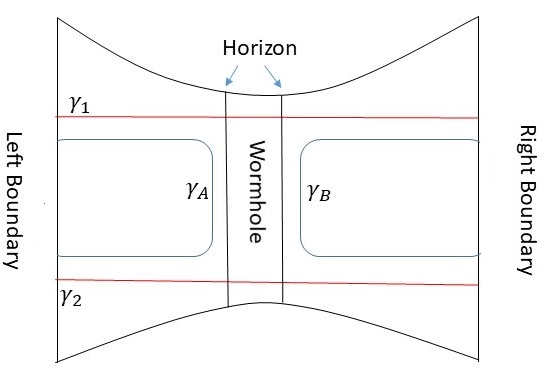}
\end{tabular}
\caption{Schematic diagram of perturbed two sided geometry}
\label{ptfd}
\end{figure}

\begin{figure}[h!]
 \centering
 \includegraphics[width=8cm,height=6cm]{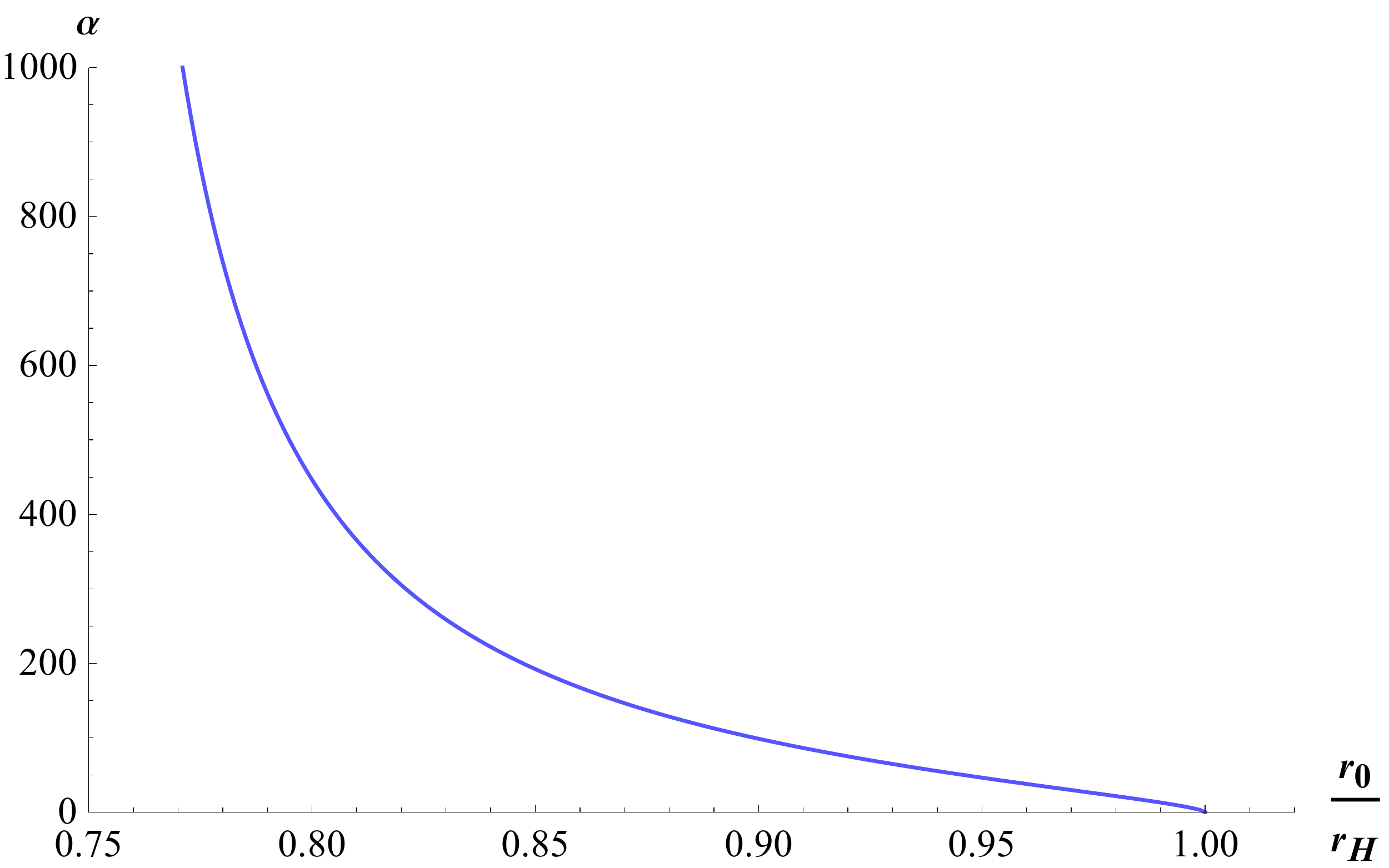}
 \caption{Variation of shockwave parameter $\alpha$ with dimensionless ratio $\frac{r_0}{r_H}$}
 \label{alphar0}
 \end{figure}

Now, We will discuss the disruption of two sided mutual information in presence of a shockwave. For the sake of simplicity we will consider the shock profile to be homogeneous. In that case the shock parameter $\alpha$ can be written as~$\alpha=\text{Constant}* e^{\frac{2\pi t_0}{\beta}}$. The mutual information in this case can be described as,
\begin{equation}
I(A,B;\alpha)=S(A)+S(B)-S(A\cup B;\alpha),
\label{Int}
\end{equation}
As the exponentially boosted shockwave modifies the near horizon region of the black hole geometry quite significantly, only $S(A\cup B)$ will receive the possible correction with increasing shockwave parameter $\alpha$. Because of the same reason, $S(A)$ and $S(B)$ will not get modified at all. To get rid of the UV divergences that appeared in $S_{A\cup B}$, it is convenient to define the following regularized version of it,  which is given as,
\begin{equation}
S^{reg}(A\cup B;\alpha)=S(A\cup B;\alpha)-S(A\cup B;\alpha=0).
\label{reg}
\end{equation}
With this regularised entanglement entropy, one can rewrite the definition of mutual information (\ref{Int}) as,
\begin{equation}
I(A,B;\alpha)=I(A,B;\alpha=0)-S^{reg}(A\cup B;\alpha).
\label{mutual}
\end{equation}
Here we will consider sufficiently large value for the strip width $l$ so as to get a finite positive result of the mutual information. In presence of the shockwave, there is an effective increase of the size of the wormhole at boundary time $t=0$ which is schematically represented in figure-\textbf{\ref{ptfd}}.
As already explained, in this case only $S(A\cup B)$ will be modified. Initially, It is given by the area of the hypersurface $\gamma_1\cup\gamma_2\equiv\gamma_{wormhole}$ and one obtains a nonzero value for the mutual information. However, as the strength of the shockwave parameter $\alpha$ increases with time, the area of the wormhole surface also increases eventually decreasing the mutual information. With further increase of $\alpha$, the area of $\gamma_{wormhole}$ at some point gets bigger than the combined area $\gamma_{A}\cup \gamma_{B}$. Hence in the definition of mutual information (\ref{mutual}), instead of $\gamma_{wormhole}$, the combined area of $\gamma_{A}$ and $\gamma_{B}$ will be favourable for the entropy $S_{A\cup B}$ due to the minimum area criteria of HRRT formalism. As a result the mutual information goes to zero, $I(A,B;\alpha)=0$.
To construct the wormhole surface we choose an appropriate embedding given as, $x^{m}=(t,0,x_2,x_3,x_4,x_5,r(t),\theta_1,\theta_2,\theta_3)$. Using this embedding one can easily obtain the components of the induced metric as,
 \begin{eqnarray}
 g_{tt}&=&G_{tt}+G_{rr}\dot{r}^2,\nonumber\\
 g_{\theta_i\theta_i}&=&G_{\theta_i\theta_i}\times~\text{ metric~ of~} S^3,
 \end{eqnarray}
The area functional can be expressed as,
 \begin{eqnarray}
\mathcal{ A}(\gamma_{wormhole})&=&2\Omega_3R^4\int dt ~r^3\sqrt{\frac{A(r)}{f(r)}}\left(-f(r)^2+A(r)\dot{r}^2\right)^{1/2},\nonumber
\\
&=&2\Omega_3R^4\int dt\mathcal{F}(r,\dot{r},t).
\label{Area}
 \end{eqnarray}
As the integrand $\mathcal{F}(r,\dot{r},t)$ in the above equation is invariant under time translation, the corresponding conserved quantity can be calculated as,
\begin{eqnarray}
 C&=&\frac{\partial \mathcal{F}}{\partial \dot{r}}\dot{r}-\mathcal{F},\nonumber\\
 &=&\frac{r^3\sqrt{A(r)}f(r)^{3/2}}{\left(-f(r)^2+A(r)\dot{r}^2\right)^{1/2}}.
 \label{Conserved}
 \end{eqnarray}
\begin{figure}[h!]
\centering
\begin{tabular}{c}
\includegraphics[width=.5\textwidth]{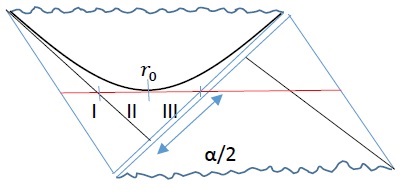}
\end{tabular}
\caption{Penrose diagram of perturbed geometry with the constant $r=r_{0}$ surface and the horizontal extremal surface as indicated by black and red lines respectively.}
\label{penrosep}
\end{figure}
Referring to figure-\textbf{\ref{penrosep}}, we note that due to the shockwave perturbation, the left half of the horizontal extremal surface will be divided into three parts with $r_{0}$ being the contact point of the constant $r$ surface and the same horizontal extremal surface.
The conserved quantity $C$ can be calculated at $r=r_0$ as,
\begin{equation}
C(r_0)=-r_0^3\sqrt{A(r_0)}\sqrt{-f(r_0)}.
\label{turn}
 \end{equation}
Next we substitute the above expression of $C(r_{0})$ into equation (\ref{Conserved}) and solve for $\dot{r}$  to get,
\begin{equation}
\dot{r}=\frac{f(C(r_{0})^{-2}r^6f(r)A(r)+1)^{1/2}}{\sqrt{A(r)}}.
 \label{extremal}
\end{equation}
Putting the above result for $\dot{r}$ in equation (\ref{Area}), we get the on shell action as,
\begin{equation}
\mathcal{A}(\gamma_{wormhole})=2\Omega_3R^4\int dr\frac{r^6 A^{3/2}}{(f A r^6C^{-2}+1)^{1/2}}.
\end{equation}
Also from the expression of $\dot{r}$ as given in (\ref{extremal}), one get the following result for time $t$ running along the wormhole as,
\begin{equation}
t=\int\frac{dr}{\dot{r}}=\int dr\frac{\sqrt{A}}{f\left(C^{-2}fAr^6+1\right)^{1/2}}.
\end{equation}
Taking all the three segments of the left half of the extremal surface in figure-\textbf{\ref{penrosep}} into consideration, the final integral form of $S_{A\cup B}$ is given as,
\begin{equation}
S(A\cup B;r_0)=\frac{\Omega_3R^4}{G^{10}_N}\Bigg(\int_{r_H}^{r_{\delta}} dr\frac{r^6 A^{3/2}}{(f A r^6C^{-2}+1)^{1/2}}+2\int_{r_0}^{r_H} dr\frac{r^6 A^{3/2}}{(f A r^6C^{-2}+1)^{1/2}}\Bigg).
\end{equation}
Here the turning point at $r=r_{0}$ also acts as the strength of the shockwave such that as the shockwave parameter $\alpha$ is increased the turning point moves further into the black hole interior and the value of $r_{0}$ decreases. Notice also that in the above equation we expressed the entanglement entropy $S_{A\cup B}$ as a function of the turning point $r_{0}$.
Finally, the corresponding regularized entropy is given as,
\begin{equation}
S^{reg}(A\cup B;r_0)=\frac{\Omega_3R^4}{G^{10}_N}\Bigg(\int_{r_H}^{r_{\delta}} dr\Big(\frac{r^6 A^{3/2}}{(f A r^6C^{-2}+1)^{1/2}}-\frac{r^3A}{\sqrt{f}}\Big)+2\int_{r_0}^{r_H} dr\frac{r^6 A^{3/2}}{(f A r^6C^{-2}+1)^{1/2}}\Bigg).
\label{entropy}
\end{equation}
Following the detailed analysis as presented in \cite{Fischler:2018kwt}, one can determine the shockwave parameter $\alpha$ in terms of the turning point $r_0$ which takes the following form,
\begin{equation}
\alpha(r_0)=2\exp(K_1(r_0)+K_2(r_0)+K_3(r_0)),
\label{sp}
\end{equation}
where $K_1,K_2$ and $K_3$ are given as follows,
\begin{eqnarray}
K_{1}(r_0)&=&\frac{4\pi}{\beta}\int_{\bar{r}}^{r_{0}}\frac{\sqrt{A}}{f} dr ,\\
K_{2}(r_0)&=&\frac{2\pi}{\beta}\int_{r_H}^{r_{\delta}}\frac{\sqrt{A}}{f}(1-\frac{1}{\sqrt{C^{-2}fAr^6+1}}) dr,\\
K_{3}(r_0)&=&\frac{4\pi}{\beta}\int_{r_0}^{r_{H}}\frac{\sqrt{A}}{f}(1-\frac{1}{\sqrt{C^{-2}fAr^6+1}}) dr.
\end{eqnarray}
Using the above we have shown the explicit dependence of $\alpha$ on the dimensionless ratio $r_0/r_h$ in figure-\textbf{\ref{alphar0}}. We observe that the shockwave parameter $\alpha$ increases with decreasing values of $r_{0}/r_{H}$ and beyond certain value $\alpha$ sharply diverges. Also as expected, at $r_{0}=r_{H}$ the shockwave parameter $\alpha$ exactly goes to zero.

Analysing the nature of the integral form of $K_3(r_0)$, we find that the sharp divergence of $\alpha$ below some critical value of $r_{0}=r_{c}$ is due to the diverging nature of $K_3(r_0)$ which can be realized from equation (\ref{sp}). To obtain this critical value, one needs to series expand the integrand of $K_{3}$ about $r=r_{0}$ and then equate the coefficient of $(r-r_{0})$ in the resulting expansion to zero \cite{Fischler:2018kwt}, yielding the following expression of $r_{c}$,
\begin{equation}
 r_c=\frac{r_H\sqrt{1-(\frac{L}{r_H})^2+\sqrt{1+(\frac{L}{r_H})^2+(\frac{L}{r_H})^4}}}{\sqrt{3}}.
\label{Critical_Radius}
\end{equation}
\subsubsection{Variation of mutual information with Shock parameter $\alpha$}
\begin{figure}[h!]
\centering
\includegraphics[width=8cm,height=5.5cm]{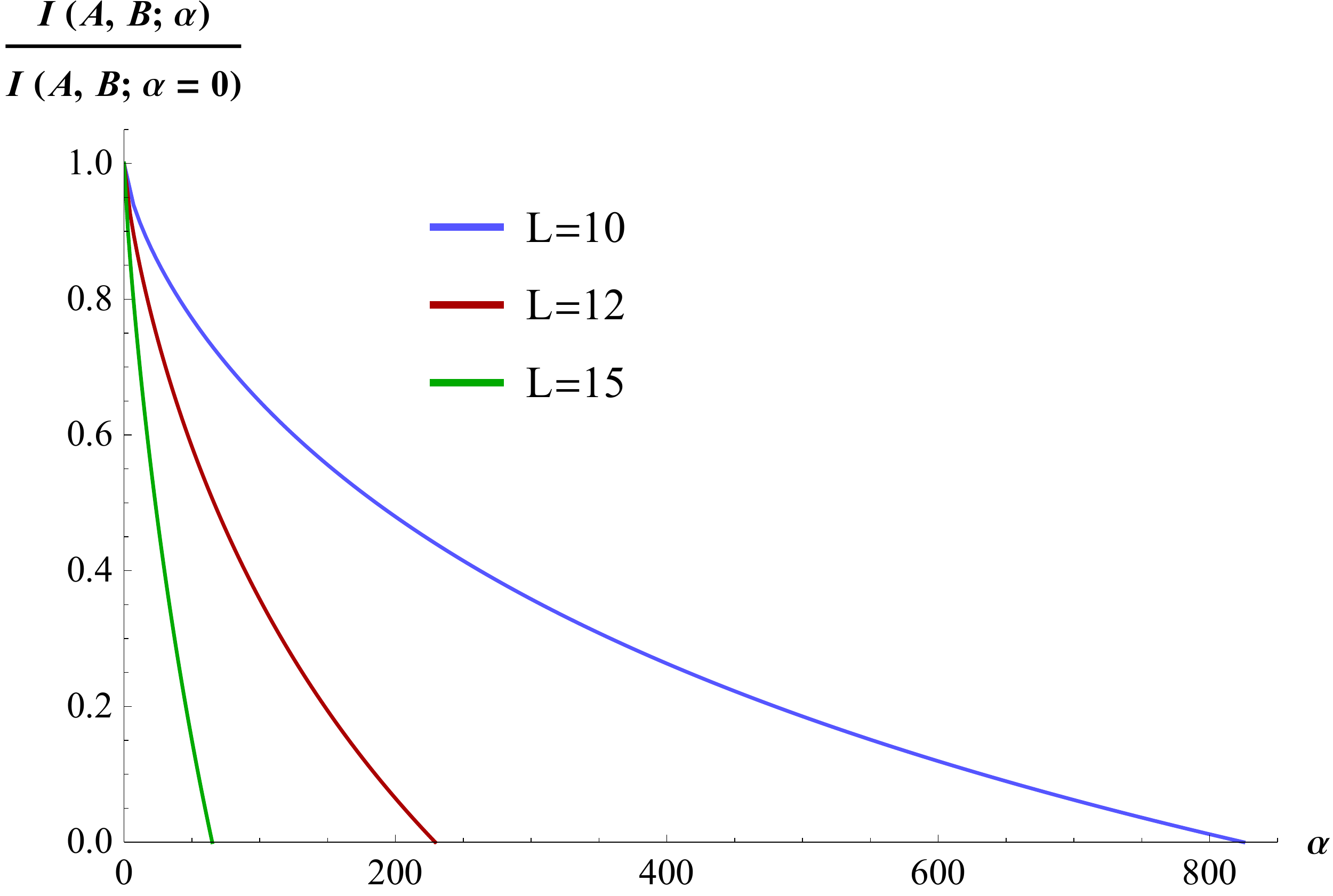}
\caption{Plot shows variation of normalised mutual inforamtion~$I(A,B:\alpha)$ with shockwave parameter $\alpha$ for different $L$}
\label{mutualalpha}
\end{figure}

Two sided mutual information in presence of shockwave is already obtained in equation (\ref{mutual}). It will be convenient to normalize it with the corresponding result at zero perturbation, that is with $\alpha=0$,
\begin{equation}
\frac{I(A,B:\alpha)}{I(A,B:\alpha=0)}=1-\frac{S^{reg}(A\cup B;r_0)}{I(A,B:\alpha=0)}.
\label{mutualf}
\end{equation}
In figure-\textbf{\ref{mutualalpha}} we have plotted the normalized mutual information as given in (\ref{mutualf}) with increasing $\alpha$ for three different values of $L$.
Form the figure we see that mutual information monotonically decreases with increasing values of $\alpha$ and eventually becomes zero. This is an expected result as the increase in shockwave parameter destroys the correlation between the left and the right side of the extended geometry. Apart from this the most important point of this analysis is the fact that for smaller length scale, for example the plot with $L=10$ (blue), the rate of decrease of mutual information with $\alpha$ is relatively slower than the one with larger value, that is $L=15$ (green). This behavior reflects the fact that the information scrambling is faster as we probe into the nonlocal regime with larger values of $L$.
\subsubsection{Entanglement velocity}
\begin{figure}[h!]
\centering
\includegraphics[width=8cm,height=6cm]{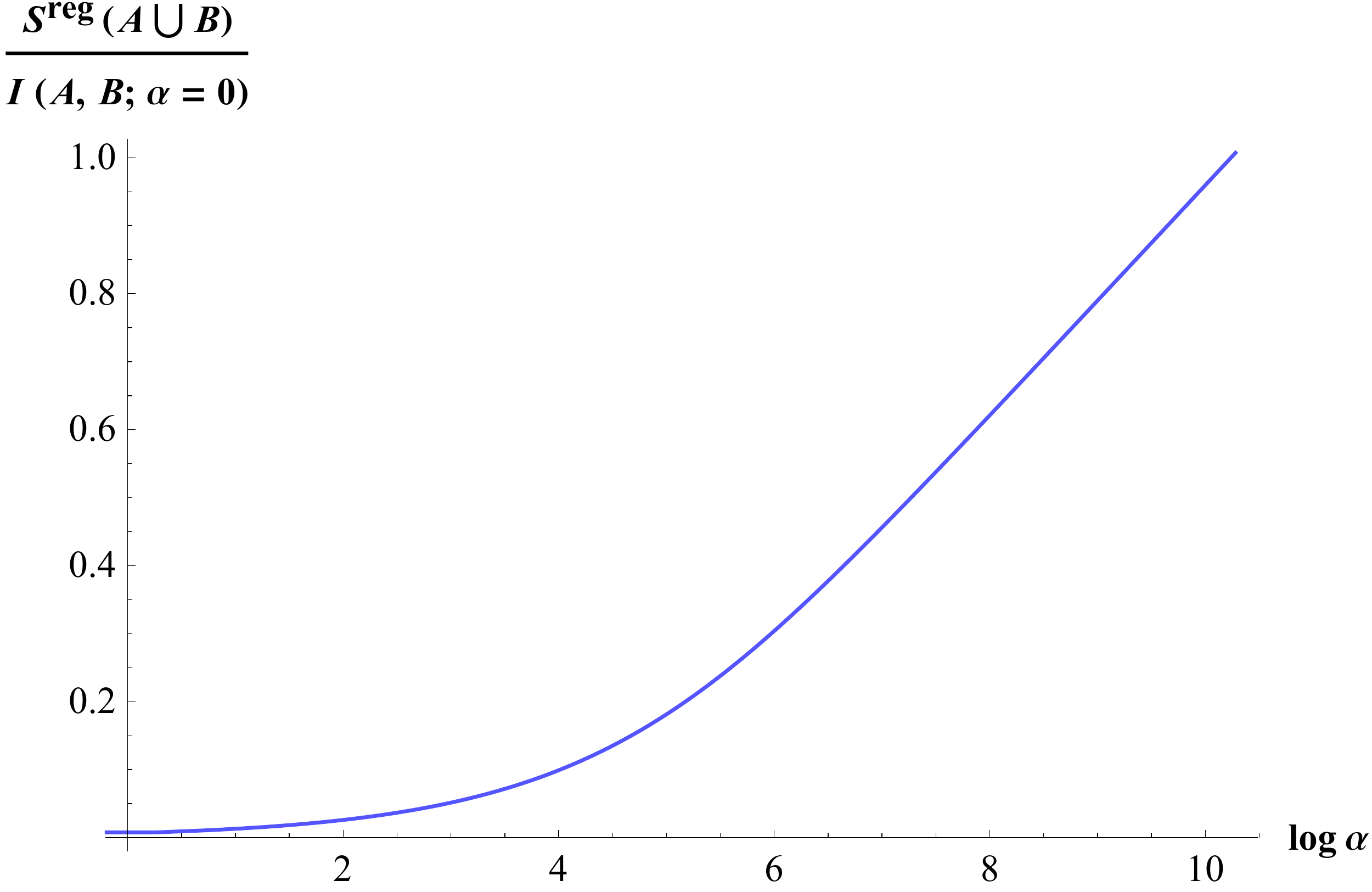}
\caption{Plot shows variation of normalised entropy $S^{reg}(A\cup B)$ with  $\text{log}~\alpha$ }
\label{regentropy}
\end{figure}

In this section we will calculate entanglement velocity $v_E$ which characterizes the spreading of entanglement pattern in chaotic systems after a global quench \cite{Avila:2018sqf,Hartman:2013qma,Liu:2013qca,Liu:2013iza,Mezei:2016zxg}. To calculate $v_E$ we need to study the variation of $S^{reg}(A\cup B;r_0)$ (\ref{entropy}) with boundary time $t_0$ at which the small perturbation is applied. We observe a linear growth of $S^{reg}(A\cup B;r_0)$ with $t_{0}$ or equivalently with $\log{\alpha}$ for homogeneous shock as shown in figure-\textbf{\ref{regentropy}} \cite{Avila:2018sqf}.
Now, one can approximate the relation between $S^{reg}(A\cup B;r_0)$ and $t_0$ in the region where it is growing linearly as,
\begin{equation}
S^{reg}(A\cup B;r_0)=\frac{R^4r_c^3}{\tilde{G}^{10}_{N}}\sqrt{-f(r_c)A(r_c)}t_0,
\end{equation}
where, $\tilde{G}^{10}_{N}=\frac{G^{10}_{N}}{\Omega_3}$ is the scaled Newton's constant of gravity in $10$-dimension. This approximation is valid only if $r_0$ is chosen in the vicinity of $r_c$ where $\alpha$ takes large values. Differentiating the above equation with respect to $t_0$, the rate of change of $S^{reg}(A\cup B)$ can be obtained as,
\begin{equation}
\frac{dS^{reg}(A\cup B;r_0)}{dt_0}=\frac{R^4r_c^3}{\tilde{G}^{10}_{N}}\sqrt{-f(r_c)A(r_c)}.
\label{velocity}
\end{equation}
Next, we calculate the thermal entropy using the Bekenstein-Hawking formula $s_{th}=\frac{\mathcal{A}(\text{horizon})}{4\tilde{G}^{10}_{N}}$ \cite{Bekenstein:1973ur}. For little string theory it can expressed in the following form,
\begin{equation}
s_{th}=\frac{\sqrt{A(r_H)}r_H^3}{4\tilde{G}^{10}_N}.
\label{thermal}
\end{equation}
Using the above expression of the thermal entropy, one can rewrite equation (\ref{velocity}) as,
\begin{equation}
\frac{dS^{reg}(A\cup B;r_0)}{dt_0}= s_{th}A_{\Sigma}\frac{r_c^3}{r_H^3}\sqrt{\frac{-f(r_c)A(r_c)}{A(r_H)}},
\label{velocity1}
\end{equation}
where $A_{\Sigma}=4R^4$ denotes the area of $\Sigma=\partial(A\cup B)$. Now comparing (\ref{velocity1}) with equation (\ref{ev}) we can obtain entanglement velocity as
\begin{equation}
v_E=\frac{r_c^3}{r_H^3}\sqrt{\frac{-f(r_c)A(r_c)}{A(r_H)}}.
\label{Entanglement_Velocity}
\end{equation}
\begin{figure}[h!]
\centering
\includegraphics[width=8cm,height=5cm]{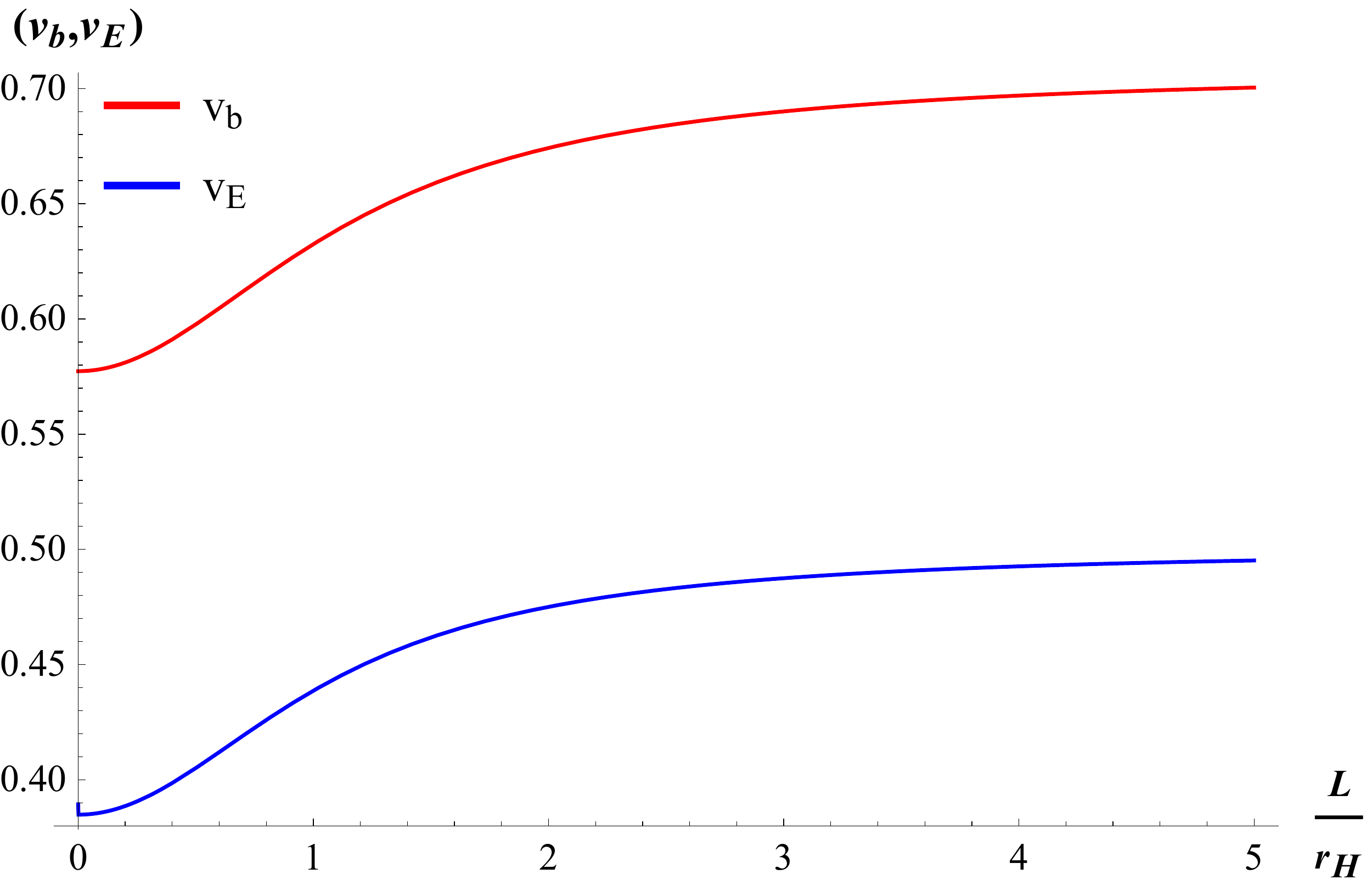}
\caption{Plot of butterfly velocity $v_B$ and entanglement velocity $v_E$ with lengthscale $\frac{L}{r_H}$  }
\label{velo}
\end{figure}
We have shown the variation of $v_b$ and $v_E$ with $L/r_H$ in figure-\textbf{\ref{velo}}. From the plot we can conclude that entanglement velocity $v_E$ and butterfly velocity $v_b$ both increases with increasing $L$. However $v_E$ always remains well below butterfly velocity $v_b$ obeying the bound as discussed in the introduction section.
\section{Pole skipping}
In this section we will explicitly compute the special locations in the complex $w-q$ plane known as the pole skipping points where the near horizon solution of bulk field equation of motion cannot be determined uniquely resulting in non-uniqueness of the corresponding retarded Green's function. Depending upon the kind of bulk field perturbations, this phenomenon can occur both in the upper half and in the lower half complex plane. For example with the components of metric perturbation in the sound channel which corresponds to the energy density retarded Green's function, the pole skipping happens in the upper half of the complex plane. Also only the upper half pole skipping points are related directly to the parameters of quantum chaos. In the following we will study the pole skipping phenomenon at lowest order in the near horizon expansion occurring on the upper half plane and compute the Lyapunov exponent and butterfly velocity. Later we will also extend the analysis for higher order poles in the lower half plane.
\subsection{Pole skipping in the upper half plane: Butterfly velocity}
In this computation we need to do a near horizon analysis of the bulk field equation of motion. Hence it is convenient to work in the ingoing coordinate system. The ten dimensional supergravity solution for the stack of NS$5$ branes in terms of the ingoing Eddington–Finkelstein coordinate $v=t+r_{*}$ is given as,
\begin{equation}\label{MetricEF}
ds^2=-f(r)A^{-1/4}dv^2+2A^{1/4}dvdr+A^{-1/4}\left(\sum_{i=1}^{5}dx_{i}^2\right)+r^2A^{3/4}d\Omega_{3}^2.
\end{equation}
Linearized perturbation of the metric and the scalar field has the following form,
\begin{equation}
\begin{split}
g_{\mu\nu}&=g_{\mu\nu}^{(0)}+h_{\mu\nu},\\
\Phi&=\Phi_{0}+\phi,
\end{split}
\end{equation}
where $g_{\mu\nu}^{(0)}$ and $\Phi_{0}$ denotes the background values for the metric and the scalar field with $h_{\mu\nu}$ and $\phi$ as the corresponding linear perturbations. The fourier transformation for the field fluctuations are given as,
\begin{equation}\label{fluc}
\begin{split}
h_{\mu\nu}(t,x_{1},r)&=e^{-i w v+i q x_{1}}h_{\mu\nu}(r),\\
\phi(t,x_{1},r)&=e^{-i w v+i q x_{1}}\phi(r),
\end{split}
\end{equation}
where we have assume the perturbation to travel along the $x_{1}$ direction. Moreover, based on the rotational symmetry in the space formed by the coordinates $x_{2},x_{3},x_{4},x_{5}$, one can categorize the metric perturbations as the following three modes,
\begin{itemize}\label{parallel}
\item Sound modes: $h_{vv},h_{vr},h_{rr},h_{vx_{1}},h_{rx_{1}}, h_{x_{1}x_{1}}, h_{x_{2}x_{2}}=h_{x_{3}x_{3}}=h_{x_{4}x_{4}}=h_{x_{5}x_{5}}$,
\item Shear modes: $h_{vx_{i}},h_{rx_{i}}, h_{x_{1}x_{2}},~~i=2,...5$,
\item Tensor mode: $h_{x_{i}x_{j}},~~i\neq j, ~~i,j=2,...5$.
\end{itemize}
We will start with the computation of butterfly velocity for which we need to consider the components of the metric perturbations in the sound channel along with the scalar field which couples to the metric. The linearized Einstein's equation is given as,
\begin{equation}\label{EOM1}
\begin{split}
&R^{(1)}_{\mu\nu}-\frac{1}{2}\left(\partial_{\mu}\Phi_{0}\partial_{\nu}\phi+\partial_{\mu}\phi\partial_{\nu}\Phi_{0}\right)+\frac{1}{4}e^{-\Phi_{0}}
\biggl\{H_{\mu b c}H_{\nu b^{\prime}c^{\prime}}\left(g^{(0)cc^{\prime}}h^{bb^{\prime}}+g^{(0)bb^{\prime}}h^{cc^{\prime}}\right)+\phi H_{\mu bc}
H_{\nu}^{bc}\biggr\}\\&
-\frac{1}{48}g^{(0)\mu\nu}e^{-\Phi_{0}}\biggl\{H_{abc}H_{a^{\prime}b^{\prime}c^{\prime}}\left(g^{(0)aa^{\prime}}g^{(0)bb^{\prime}}h^{cc^{\prime}}
+g^{(0)bb^{\prime}}g^{(0)cc^{\prime}}h^{aa^{\prime}}+g^{(0)cc^{\prime}}g^{(0)cc^{\prime}}h^{bb^{\prime}}\right)-\phi H_{3}^{2}\biggr\}\\&
+\frac{1}{48}e^{-\Phi_{0}}H_{3}^2 h_{\mu\nu}=0.
\end{split}
\end{equation}
Also the linearized scalar field equation of motion is given as,
\begin{equation}\label{EOM2}
\begin{split}
&g^{(0)\mu\nu}\left(\partial_{\mu}\partial_{\nu}\phi-\Gamma^{(1)\rho}_{\mu\nu}\partial_{\rho}\Phi_{0}-\Gamma^{(0)\rho}_{\mu\nu}\partial_{\rho}\phi
\right)+h^{\mu\nu}\left(\partial_{\mu}\partial_{\nu}\Phi_{0}-\Gamma^{(0)\rho}_{\mu\nu}\partial_{\rho}\Phi_{0}\right)-\frac{1}{12}e^{-\Phi_{0}}\phi H_{3}^2
\\& -\frac{1}{12}e^{-\Phi_{0}}H_{abc}H_{a^{\prime}b^{\prime}c^{\prime}}\left(g^{(0)aa^{\prime}}g^{(0)bb^{\prime}}h^{cc^{\prime}}
+g^{(0)bb^{\prime}}g^{(0)cc^{\prime}}h^{aa^{\prime}}+g^{(0)cc^{\prime}}g^{(0)cc^{\prime}}h^{bb^{\prime}}\right)=0,
\end{split}
\end{equation}
where, $\Gamma^{(0)\rho}_{\mu\nu}$ is the affine connection corresponding to the background value of the metric while $\Gamma^{(1)\rho}_{\mu\nu}$, $R^{(1)}_{\mu\nu}$ are the linearized fluctuations to the affine connection and the ricci tensor respectively defined as,
\begin{equation}\label{ricci}
\begin{split}
\Gamma^{(0)\rho}_{\mu\nu}&=\frac{1}{2}g^{(0)\lambda\rho}\left(\partial_{\mu}g^{(0)}_{\lambda\nu}+\partial_{\nu}g^{(0)}_{\lambda\mu}
-\partial_{\lambda}g^{(0)}_{\mu\nu}\right),\\
\Gamma^{(1)\rho}_{\mu\nu}&=\frac{1}{2}\left[g^{(0)\lambda\rho}\left(\partial_{\mu}h_{\lambda\nu}+\partial_{\nu}h_{\lambda\mu}
-\partial_{\lambda}h_{\mu\nu}\right)-h^{\lambda\rho}\left(\partial_{\mu}g^{(0)}_{\lambda\nu}+\partial_{\nu}g^{(0)}_{\lambda\mu}
-\partial_{\lambda}g^{(0)}_{\mu\nu}\right)\right],\\
R^{(1)}_{\mu\nu}&=\partial_{\rho}\Gamma^{(1)\rho}_{\mu\nu}-\partial_{\mu}\Gamma^{(1)\rho}_{\rho\nu}
+\Gamma^{(0)\rho}_{\rho\lambda}\Gamma^{(1)\lambda}_{\mu\nu}+\Gamma^{(1)\rho}_{\rho\lambda}\Gamma^{(0)\lambda}_{\mu\nu}
-\Gamma^{(0)\rho}_{\mu\lambda}\Gamma^{(1)\lambda}_{\rho\nu}-\Gamma^{(1)\rho}_{\mu\lambda}\Gamma^{(0)\lambda}_{\rho\nu}.
\end{split}
\end{equation}
In this case the non-uniqueness of retarded Green's function is due to the fact that the $vv$ component of the linearized Einstein's equation (\ref{EOM1}) becomes trivial at the pole skipping point. Hence in the near horizon limit we first expand the field perturbations as,
\begin{equation}
\begin{split}
h_{\mu\nu}&=\h^{(0)}_{\mu\nu}+\h^{(1)}_{\mu\nu}\left(r-r_{H}\right)+\mathcal{O}(r-r_{H})^2,\\
\phi&=\phi^{(0)}+\phi^{(1)}\left(r-r_{H}\right)+\mathcal{O}(r-r_{H})^2.
\end{split}
\end{equation}
Substituting the above expansion in the Einstein's equation and expanding the same near the horizon we obtain,
\begin{equation}\label{EOM3}
\begin{split}
&h^{(0)}_{vv}\biggl\{L^2\left(Q^2-2i W\right)+r_{H}^2\left(Q^2-3i W\right)\biggr\}\\&+\left(W-i\right)\left(L^2+r_{H}^2\right)\biggl\{2Q h_{vx_{1}}^{(0)}+W\left(h_{x_{1}x_{1}}^{(0)}+4h_{x_{2}x_{2}}^{(0)}\right)\biggr\}=0,
\end{split}
\end{equation}
where we define $W=w/2\pi T_{H}$ and $Q=q/2\pi T_{H}$, The above equation is satisfied for the particular values of $W$ and $Q$ given as,
\begin{equation}
W=i,~~~~Q=i\sqrt{\frac{2L^2+3r_{H}^2}{L^2+r_{H}^2}}.
\end{equation}
The Lyapunov exponent and the butterfly velocity is given as,
\begin{equation}
\lambda_{L}=2\pi T_{H},~~~~
v_{B}^2=\frac{L^2+r_{H}^2}{2L^2+3r_{H}^2}.
\end{equation}
\subsection{Pole-skipping at higher order complex frequencies}
In the last subsection we discussed the pole skipping point at the lowest order in the upper half plane and obtained the expression of butterfly velocity. In this section we will extend the analysis of pole skipping phenomenon for higher ordered complex
frequencies $W_{n}$, with $n=1,2,3,...$.
\subsubsection{Scalar field fluctuations}
Considering linearized fluctuation of the scalar field as given in (\ref{fluc}), the corresponding equation of motion can be obtained as,
\begin{equation}\label{scalareq}
\phi^{\prime\prime}(r)+\mathcal{S}(r)\phi^{\prime}(r)+\mathcal{R}(r)\phi(r)=0,
\end{equation}
where $\mathcal{S}(r)$ and $\mathcal{R}(r)$ are given as,
\begin{equation}\label{SR}
\begin{split}
\mathcal{S}(r)&=\frac{1}{r\left(r^2-r_{H}^2\right)}\biggl(-3r^2+r_{H}^2+2iWr^2\sqrt{\frac{L^2+r^2}{L^2+r_{H}^2}}\biggr),\\
\mathcal{R}(r)&=\frac{1}{r^2-r_{H}^2}\Biggl(2L^2\frac{L^2+r_{H}^2}{\left(L^2+r^2\right)^2}+Q^2\left(\frac{L^2+r^2}{L^2+r_{H}^2}\right)
+iW\frac{\left(2L^2+3r^2\right)}{\sqrt{L^2+r^2}\sqrt{L^2+r_{H}^2}}\Biggr).
\end{split}
\end{equation}
To determine the pole skipping points at different orders one needs to consider the near horizon expansion for the solution $\phi(r)$ of the above equation as,
\begin{equation}
\phi(r)=\phi_{0}+\left(r-r_{H}\right)\phi_{1}+\left(r-r_{H}\right)^2\phi_{2}+\left(r-r_{H}\right)^3\phi_{3}+...
\end{equation}
Substituting the above into equation (\ref{scalareq}), and equating the coefficients of $\left(r-r_{H}\right)^n$ to zero for $n=0,1,2,3,..$ one gets the following system of linear equations,
\begin{equation}\label{eqnorder}
\begin{split}
c_{00}\phi_{0}+c_{01}\phi_{1}&=0,\\
c_{10}\phi_{0}+c_{11}\phi_{1}+c_{12}\phi_{2}&=0,\\
c_{20}\phi_{0}+c_{21}\phi_{1}+c_{22}z_{2}+c_{23}\phi_{3}&=0,\\
~~~~~~~~~~~~~~~~ \vdots
\end{split}
\end{equation}
where the coefficients $c_{ij}$'s can be arranged in a square matrix of a given order as,
\begin{equation}
\label{Cmatrix}
C= \begin{pmatrix}
	c_{00} && c_{01} && 0 && 0 && \cdots\\
	c_{10} && c_{11} && c_{12} && 0 && \cdots\\
	c_{20} && c_{21} && c_{22} && c_{23} && \cdots\\
	\vdots && \vdots && \vdots && \vdots && \cdots \\,
\end{pmatrix},
\end{equation}
such that the pole skipping points are obtained by solving simultaneously the following two equations,
\begin{equation}\label{matrixeqn}
c_{n-1~n}=0,~~~~~~\det{C}=0.
\end{equation}
In this case with the large $L$ limit, the first few pole skipping points are given as,
\begin{equation}\begin{split}
W_1&=-i~~~~~~Q_{1}^2=-2+2\frac{r_{H}^2}{L^2}+\mathcal{O}\biggl(\frac{r_{H}^4}{L^4}\biggr)\\
W_{2}&=-2i~~~~~Q_{2}^2=\left\{-4-i2\sqrt{6}\frac{r_{H}}{L},-4+i2\sqrt{6}\frac{r_{H}}{L}\right\}+\mathcal{O}\biggl(\frac{r_{H}^2}{L^2}\biggr)\\
W_{3}&=-3i~~~~~Q_{3}^2=\left\{-6-\left(12+i6\sqrt{3}\right)\frac{r_{H}^2}{L^2},-6-\left(12-i6\sqrt{3}\right)\frac{r_{H}^2}{L^2},
-10+22\frac{r_{H}^2}{L^2}\right\}
+\mathcal{O}\biggl(\frac{r_{H}^4}{L^4}\biggr)
\end{split}
\end{equation}
\subsubsection{Shear mode of metric fluctuations}
The gauge invariant variable for the shear modes of metric perturbations is given as,
\begin{equation}
Z_{2}(u)=Q H_{v x_{2}}(u)+W H_{x_{1}x_{2}}(u),
\end{equation}
where we have considered the redial gauge such that $h_{\mu u}=0$ for any $\mu$. In the above equation the field fluctuations are defined as
$H_{x_{1}x_{2}}=h_{x_{1}x_{2}}/g_{x_{1}x_{1}}$ and $H_{v x_{2}}=h_{v x_{2}}/g_{x_{2}x_{2}}$.
The Einstein's equation of motion for the shear modes of field perturbation is obtained in terms of the gauge invariant variable $Z_{2}$ as,
\begin{equation}
Z_{2}^{\prime\prime}(u)-\frac{W^2}{\left(1-u\right)\left(W^2-(1-u)Q^2\right)}Z_{2}^{\prime}(u)+\frac{\left(r_{H}^2+u L^2\right)\left(W^2-(1-u)Q^2\right)}{4u^3\left(1-u\right)^2\left(r_{H}^2+L^2\right)}Z_{2}(u).
\end{equation}
Using the method as discussed previously we obtain the pole skipping points for the first
few orders in the limit $\left(r_{H}/L\right)\rightarrow 0$ as,
\begin{equation}\begin{split}
W_1&=-i~~~~~~Q_{1}^2=2-2\frac{r_{H}^2}{L^2}+\mathcal{O}\biggl(\frac{r_{H}^4}{L^4}\biggr)\\
W_{2}&=-2i~~~~~Q_{2}^2=\left\{4-\frac{r_{H}^2}{L^2},-4-5\frac{r_{H}^2}{L^2}\right\}+\mathcal{O}\biggl(\frac{r_{H}^4}{L^4}\biggr)\\
W_{3}&=-3i~~~~~Q_{3}^2=\left\{6+3\frac{r_{H}^2}{L^2},-6-3\frac{r_{H}^2}{L^2},-10-5\frac{r_{H}^2}{L^2}\right\}
+\mathcal{O}\biggl(\frac{r_{H}^4}{L^4}\biggr)
\end{split}
\end{equation}
\section{Conclusion}
In this paper we have used different well known holographic techniques to study quantum information scrambling and quantum chaos in a nonlocal theory known as the little string theory. The fundamental degrees of freedom of this theory are strings with finite length $l_{s}$. The nonlocal effects appear only at length scale lower than the inverse Hagedorn temperature $\beta_{h}$ which is proportional to the square root of the string length. The main conclusion of this analysis is that the nonlocality accelerates the scrambling of quantum information. In other words the theory becomes more and more chaotic as one probe into length scale less than the inverse Hagedorn temperature. This can be explicitly realized from the variation of butterfly velocity $v_{b}$ with $\beta_{h}=2\pi L$ as already discussed. Butterfly velocity acts as an effective bound at low energy on the spread of quantum information. An increasing behavior of $v_{b}$ with the nonlocal scale basically widen the effective butterfly cone resulting in faster growth of local operators. However the butterfly velocity is observed to saturate to a value less than the speed of light due to the Lorentz invariance.

For the holographic computation we work in the usual thermofield double set up whose dual description is given in terms of the extended two sided black hole geometry. A small perturbation on the boundary correspond to an energy quanta in the bulk which gets exponentially boosted with time and produces a shockwave. This highly energetic quanta can disrupt the entanglement structure between the left and the right side of the geometry which is actually the butterfly effect for this holographic set up. This disruption of entangled structure is captured by the decrease of holographic two sided mutual information with increasing shockwave parameter. In this case we also observe that the rate of disruption of mutual information is faster as the nonlocal scale is increased.

Regarding the chaotic behavior, we considered the recently observed phenomenon called pole skipping that provides an efficient and easier way of computing Lyapunov exponent and also butterfly velocity which characterizes quantum chaos in many body system. Analysing linearized Einstein's equation involving different bulk field perturbations near the black hole horizon singles out special quasinormal frequency and momentum in the complex plane which can be connected directly to the chaos parameters due to the phenomenon of pole skipping. In this paper we have computed the pole skipping points in the upper half plane to obtain the butterfly velocity which exactly matches with the result as obtained through the gravitational shockwave analysis in section-{\bf 5}.

In future, it will be interesting to repeat the same exercise for the gravity dual of $\mathcal{N}=4$ SYM theory which is deformed by the presence of external heavy quarks \cite{Chakrabortty:2011sp, Chakrabortty:2020ptb, Chakrabortty:2022kvq}. Further in \cite{Boruch:2020wbe}, the author has studied the dynamics of entanglement wedge cross section in shockwave geometry for three dimensional $\text{AdS}$ geometry. A generalization of this result in higher dimensional black hole background might be an useful future direction.
\section*{Acknowledgements}
We would like thank Chandrasekhar Bhamidipati for helpful discussion and comments on the final draft of this paper. KS would like to thank Shankhadeep Chakrabortty, Hironori Hoshino and especially Sanjay Pant for useful regular discussions. KS is supported by the institute post doctoral fellowship from IIT Bhubaneswar.
{}
\end{document}